\documentclass[pdflatex,sn-nature]{sn-jnl}

\usepackage{graphicx}
\usepackage{multirow}
\usepackage{amsmath,amssymb,amsfonts}
\usepackage{amsthm}
\usepackage{mathrsfs}
\usepackage[title]{appendix}
\usepackage{xcolor}
\usepackage{textcomp}
\usepackage{manyfoot}
\usepackage{booktabs}
\usepackage{algorithm}
\usepackage{algorithmicx}
\usepackage{algpseudocode}
\usepackage{braket}
\usepackage{listings}
\usepackage{lineno}
\usepackage[normalem]{ulem}

\title{Spatial nonlocality imaging via metasurface}

\author[1]{Jian Li}
\equalcont{These authors contributed equally to this work.}

\author[2]{Zi-Mu Fan}
\equalcont{These authors contributed equally to this work.}

\author[1]{Qing-Yuan Wu}
\equalcont{These authors contributed equally to this work.}

\author*[1]{Wen-Kai Yu}
\email{yuwenkai@bit.edu.cn}

\author[3]{Zhe Meng}

\author[1]{Xing-Yan Fan}

\author[1]{Wen-Hao Wang}

\author[1]{Jie Ma}

\author*[2]{Xia Guo}
\email{guox@bupt.edu.cn} 
\author*[1]{An-Ning Zhang}
\email{anningzhang@bit.edu.cn}

\affil[1]{Center for Quantum Technology Research, School of Physics, Beijing Institute of Technology, Beijing 102488, China}

\affil[2]{School of Electronic Engineering, Beijing University of Posts and Telecommunications, Beijing 100876, China}
\affil[3]{National Innovation Institute of Defense Technology, AMS, Beijing 100071, China}
\date{}

\begin{document}
\abstract{Bell nonlocality is both a defining signature of entanglement and a key quantum information resource. However, visualizing and certifying nonlocal correlations across a spatially multimode photonic field remains challenging due to the rapidly growing measurement cost of spatially resolved projective tests. To address this issue, we build a spatial nonlocality imaging scheme that directly reveals the spatial distribution of quantum nonlocality by integrating a metasurface that performs parallel polarization projections with a quantum-adaptive neural network. Spatially resolved Clauser--Horne--Shimony--Holt (CHSH) tests are realized over a 400-pixel biphoton field using an average of only 1.7 detected coincidence pairs per pixel per basis. This approach yields a nonlocality image that maps the two-dimensional spatial distribution of Bell violations across the optical field and reveals the target-state-dependent spatial evolution of Bell violations. It provides a highly resource-efficient route to large-scale Bell certification and opens new possibilities for exploiting spatially multimode entanglement in quantum imaging, quantum networking, and scalable photonic quantum technologies.}
\maketitle

\section*{Introduction}
Photonic systems have emerged as a powerful platform for scalable quantum information tasks \cite{zhongQuantumComputationalAdvantage2020, flaminiPhotonicQuantumInformation2019}, owing to their versatility, large encoding capacity, and compatibility with ready-made optical infrastructure. High-dimensional biphoton entanglement \cite{cozzolinoHighDimensionalQuantumCommunication2019,zhangEngineeringTwophotonHighdimensional2016,yu2026experimental}, encoded in spatial, temporal, or angular degrees of freedom, provides access to many parallel modes, enabling major advances in quantum communication \cite{zengControlledEntanglementSource2023, caoDirectCounterfactualCommunication2017}, sensing \cite{barreiroBeatingChannelCapacity2008a}, and imaging \cite{moreauImagingBelltypeNonlocal2019, moreauImagingQuantumStates2019, liMetalensarraybasedHighdimensionalMultiphoton2020, yangQuantumMetasurfaceHolography2022, cuiQuantumImagingExploiting2023}. Spatially distributed, multi-pixel biphoton fields \cite{goelQuantumInformationProcessing2025} are especially attractive, as they offer a natural route to massively parallel quantum measurements and high-throughput quantum imaging \cite{defiennePolarizationEntanglementenabledQuantum2021, ziaInterferometricImagingAmplitude2023}. 

However, exploiting multi-mode entanglement spatially remains hindered by the difficulty of certifying nonlocal correlations across many spatial modes. Bell-type nonlocality tests, such as the Clauser--Horne--Shimony--Holt (CHSH) inequality, provide a stringent and meaningful benchmark of quantum correlations, and their violation directly rules out any classical description based on local realism. For a single effective mode, the evaluation of the CHSH inequality traditionally requires multiple distinct joint projective measurement settings \cite{freedmanExperimentalTestLocal1972b, aspectExperimentalTestsRealistic1981, aspectExperimentalRealizationEinsteinPodolskyRosenBohm1982, collinsBellInequalitiesArbitrarily2002, turaDetectingNonlocalityManybody2014}. 
Quantum imaging has proven highly effective for the sensing and characterization of high-dimensional quantum states and spatially structured quantum fields~\cite{jackHolographicGhostImaging2009,dadaExperimentalHighdimensionalTwophoton2011,moreauImagingBelltypeNonlocal2019, ziaInterferometricImagingAmplitude2023, dongImagingBasedQuantumTomography2023, qiuRemoteTransportHighdimensional2023a, xuHybridEntanglementCarrying2025, genoveseRealApplicationsQuantum2016,chenImagingSinglePhotonOrbitalAngularMomentum2022,engineerCertifyingHighdimensionalQuantum2025}. By contrast, biphoton nonlocality measurements still rely mainly on global or mode-selective projections, lacking the spatial resolution needed for direct imaging of nonlocality. Bell nonlocality should therefore be regarded not merely as a global scalar quantity, but as a spatially distributed observable whose faithful characterization fundamentally depends on imaging resolution and measurement strategy. A fully spatially resolved Bell test requires repeating all joint measurement settings for every spatial pixel, causing the measurement resource cost to scale quadratically with the number of spatial modes. Given the limited brightness of entangled-photon sources, this scaling poses a fundamental scalability bottleneck for large-scale multi-pixel entanglement quantum technologies.

Here, we propose spatial nonlocality imaging, which enables spatially resolved imaging of Bell nonlocality across an entangled field. On the hardware side, we employ a single-layer metasurface to parallelize local measurements. On the algorithmic side, inspired by recent advances in photon-efficient imaging under photon-starved conditions \cite{kirmaniFirstPhotonImaging2014,liuFastFirstphotonGhost2018a,pengFirstphotonImagingHybrid2020,shinPhotonefficientImagingSinglephoton2016a,shinPhotonEfficientComputational3D2015}, we develop a deep-learning framework that reconstructs spatially resolved Bell nonlocality directly from only a few photon-detection pairs. Our simulations indicate that higher spatial resolution enables stronger and clearer nonlocal correlations by better preserving distinct mode separation and accurately revealing the underlying quantum correlations. Furthermore, based on experimentally generated polarization--orbital-angular-momentum (OAM) hybrid entangled states, we investigate the target-state dependence of the reconstructed nonlocality images. We further investigate how mismatches between the target and detected states affect the imaging results, highlighting the intrinsically measurement-dependent nature of CHSH nonlocality measurements. Experimentally, we verified spatially resolved CHSH tests across a 400-pixel entangled biphoton field, using only $\sim$1.7 detected photon pairs per pixel per measurement, $\sim$$10^4$ coincidence pairs in total, showing a dramatic reduction in photon-resource requirements for large-scale Bell certification. We further demonstrate two different CHSH measurement settings, which produce distinct spatial nonlocality distributions, directly illustrating the measurement-dependent nature of nonlocality imaging. This scheme establishes a resource-efficient framework for spatially resolved imaging and characterization of Bell nonlocality, offering a practical route to scalable multi-mode entanglement verification for quantum information.

\section*{Results}
\subsection*{Spatially resolved quantum nonlocality}
High-dimensional entangled quantum fields often exhibit nontrivial spatial structure, with quantum correlations distributed across the transverse plane. To provide a general and implementation-independent description, we consider a bipartite entangled state whose local quantum properties vary with the transverse position $r$, characterized by a local density operator $\rho_r$ encoding the accessible quantum correlations at each point. 

In an idealized scenario with infinite spatial resolution, nonlocal correlations could be probed at each spatial position. In practice, insufficient spatial resolution causes each pixel to integrate quantum signals over a finite spatial region. Consequently, the measured quantum state in a given pixel is not exactly $\rho_r$ but the spatial average over this pixel area, which can be expressed in the general form
\begin{equation}\label{eq1}
\rho_d=\int_{r\in d}\eta_r\rho_r\, \mathrm{d}r,
\end{equation}
where $\rho_r$ is the local density matrix at position $r$ and $\eta_r$ denotes the detection weighting function over this pixel area. Thus, finite spatial resolution inherently mixes locally structured entangled states $\rho_r$, turning them into effective mixed states. Accurately probing spatially structured quantum correlations thus requires high spatial resolution, highlighting the necessity of spatially resolved imaging for revealing nonlocality in high-dimensional entangled fields. 

Bell nonlocality provides a stringent and operationally meaningful criterion for assessing such quantum correlations. For bipartite systems, the CHSH inequality sets the classical bound $S\le2$ for any local hidden-variable model, where $S$ is defined in terms of expectation values of local observables; its violation signifies genuine quantum nonlocality. Under finite spatial resolution, the measured Bell parameter $S$ is determined by the detected mixed state $\rho_d$ and takes the form of
\begin{equation}\label{eq2}
S_d = \mathrm{Tr}(\hat{S}_{t} \cdot \rho_d)
\end{equation}
where $S_d$ is the CHSH Bell parameter detectable by the pixel, and $\hat{S}_{t}$ denotes a CHSH measurement operator constructed based on the target state $\rho_t$ (see Supplementary Materials~S1 for details). The discrepancy between the detected state $\rho_d$ and the target state $\rho_t$ can lead to distortions in the nonlocality imaging results. Such distortions mainly originate from two aspects. One is the degradation of $\rho_d$ caused by insufficient imaging resolution, which prevents the detected state from faithfully reflecting the underlying local quantum state $\rho_r$. The other arises from mismatches between $\rho_t$ and the underlying local quantum state. In the following, we first discuss the blurring of nonlocality imaging induced by insufficient spatial resolution, and then, based on our experimental framework, further analyze the effects arising from the mismatch between $\rho_t$ and $\rho_r$. 

\begin{figure}[t]
    \centering
     \includegraphics[width=0.8\textwidth]{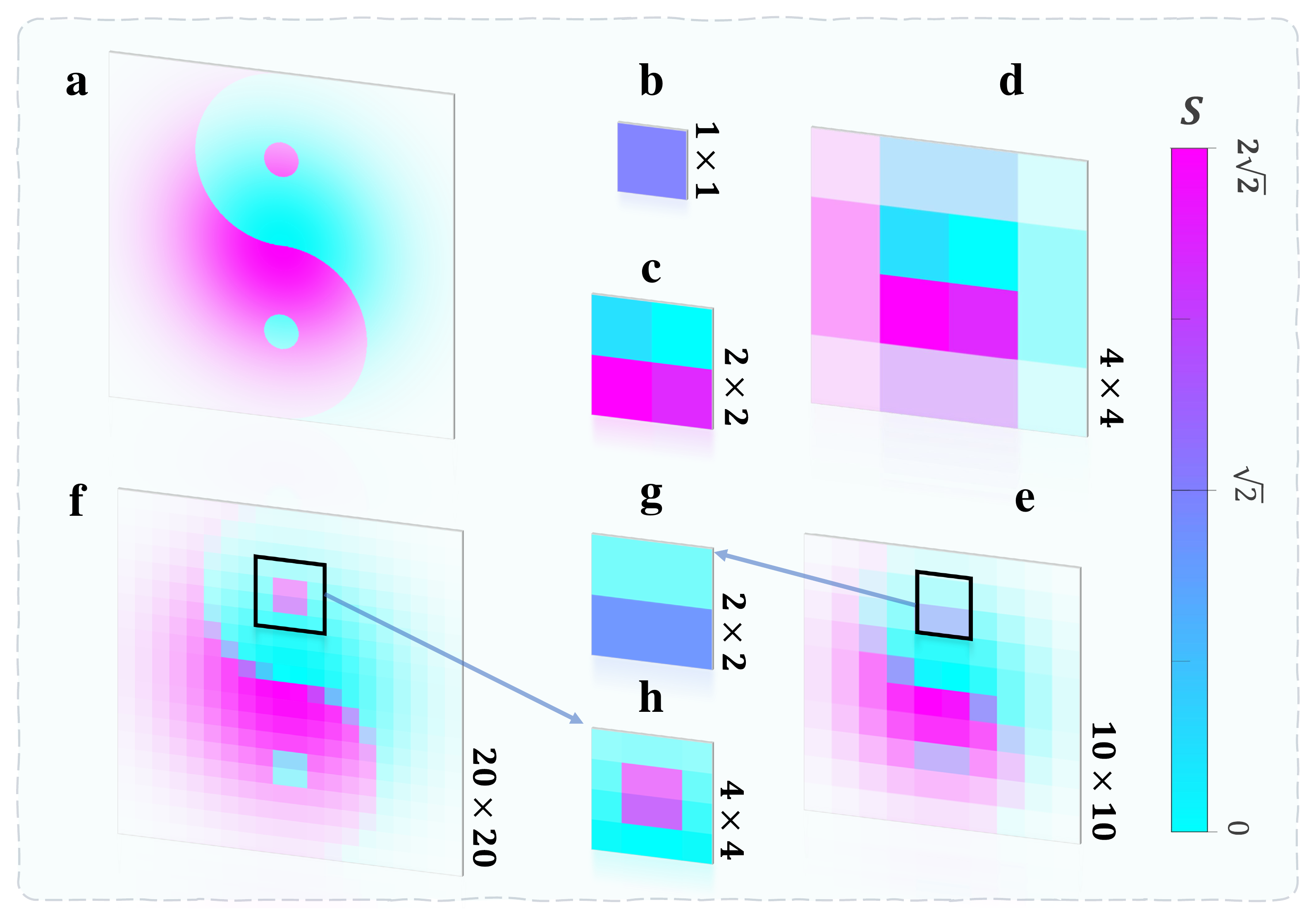} 
    \caption{\textbf{Spatial-resolution–dependent manifestation of quantum nonlocality.} \textbf{a}, simulated bipartite entangled field distribution resembling the Taiji-like spatial pattern, where blue regions show no detectable Bell violation ($S=0$), magenta regions exhibit Bell violation ($S=2\sqrt{2}>2$). \textbf{b}, result of single-mode detection (i.e., complete spatial averaging over the transverse field) yields $S_d=1.41$, below the classical bound, for which no Bell violation is observed. \textbf{c–f}, reconstructed images at increasing spatial resolution: 2$\times$2, 4$\times$4, 10$\times$10, and 20$\times$20, respectively. Finer resolution progressively resolves distinct spatial modes and recovers the spatial nonlocality distribution. \textbf{g}, \textbf{h}, magnified views of the upper magenta regions from \textbf{e} and \textbf{f}, respectively. \textbf{g}, coarse resolution gives $S_d^{max}=1.21<2$ (no violation). \textbf{h}, sufficient resolution yields $S_d^{max}=2.47>2$ (pronounced violation). These simulations show that quantum nonlocality in spatially structured entangled states is intrinsically resolution-dependent, and high spatial resolution is essential for accurately resolving spatially distributed nonlocal correlations.}
    \label{fig:1}
\end{figure}

Equations~(\ref{eq1}) and (\ref{eq2}) capture a general mechanism by which spatial resolution fundamentally limits observable Bell nonlocality. As illustrated in Fig.~\ref{fig:1}a, we consider a spatially structured biphoton polarization‑entangled state whose nonlocal correlations are distributed across the transverse plane. The spatial pattern is chosen to resemble the traditional Taiji-like spatial pattern: magenta regions violate the CHSH inequality ($S=2\sqrt{2}>2$) while blue regions indicate no detectable Bell violation ($S=0$). Under insufficient spatial resolution, distinct spatial modes are mixed via spatial averaging, progressively reducing the measured Bell parameter even without decoherence or technical noise. In the extreme case of a single effective pixel covering the entire field, the resulting mixed state no longer violates the Bell inequality, rendering nonlocality unobservable (see Fig.~\ref{fig:1}b).

Conversely, as the spatial resolution increases, detector pixels approach the characteristic length scale of the spatial variation of $\rho_r$, making distinct spatial modes well separated. As shown in Figs.~\ref{fig:1}c--\ref{fig:1}e, increasing the number of pixels enables a progressively more faithful reconstruction of the Bell nonlocality distribution. In this regime, the detected state $\rho_d$ then more closely approximates the underlying local quantum state, leading to stronger and clearer Bell violations. This effect is highlighted in the magnified views of the same region in Figs.~\ref{fig:1}g (coarse-grained detection) and \ref{fig:1}h (fine-resolution detection): coarse detection yields no violation ($S_d^{max}=1.21<2$), while fine resolution restores a clear violation ($S_d^{max}=2.47>2$). Thus, the absence of Bell violation under coarse-grained detection may arise from insufficient spatial resolution rather than the absence of nonlocality. High spatial resolution is therefore not merely a technical refinement but a fundamental requirement for faithfully characterizing spatially structured nonlocal correlations in entangled quantum fields.

\subsection*{Experimental implementation utilizing metasurface-based parallel Pauli projections}
Bell-type and CHSH nonlocality have been demonstrated in increasingly sophisticated experiments, from satellite-based implementations (e.g., Mozi) to high-efficiency optical Bell tests \cite{cabelloLoopholeFreeBellTest2012, christensenDetectionLoopholeFreeTestQuantum2013, giustinaSignificantLoopholeFreeTestBells2015, larssonLoopholesBellInequality2014, yinSatellitebasedEntanglementDistribution2017a}. These experiments typically rely on $10^3$--$10^4$ detection events to achieve precise statistical verification of Bell violations within a few well-defined measurement channels, and are intrinsically non-imaging paradigms, as they extract a single global Bell parameter from carefully selected modes. Recent studies \cite{moreauImagingBelltypeNonlocal2019, moreauImagingQuantumStates2019} have introduced imaging or spatially resolved detection as an auxiliary tool for assessing nonlocal correlations, but the spatial degree of freedom is used primarily to access or average over a small number of effective modes, rather than to reconstruct spatially distributed nonlocal correlations. Meanwhile, metasurfaces have emerged as a compact, versatile platform for quantum photonics, enabling precise subwavelength control, with demonstrated applications in quantum state acquisition, multiphoton interference, and entanglement engineering \cite{wangQuantumMetasurfaceMultiphoton2018, stavQuantumEntanglementSpin2018, santiago-cruzResonantMetasurfacesGenerating2022, gaoInterferenceInducedEntanglementEngineering2026, jiaPolarizationentangledBellState2025,yangQuantumMetasurfaceHolography2022,wuQuantumProcessTomography2022}. Notably, recent advances in polarization-dependent wavefront control and spatial multiplexing open up the possibility of applying metasurfaces to spatially resolved polarization analysis. 

While high spatial resolution is essential for revealing distributed nonlocal correlations, extending Bell tests to a multi-pixel imaging regime incurs a severe quantum resource cost. For an entangled field consisting of $N$ spatial modes, its fully resolved nonlocality reconstruction requires $16N$ projection configurations. Since the total photon flux is distributed across all spatial channels, the integration time required for each projection scales linearly with $N$, making the total acquisition time scale as $16N^2$. For a modest spatial resolution of $N=400$ pixels, this corresponds to an acquisition time exceeding $2.56\times 10^6$ times that required for a single polarization projection. Such unfavorable scaling has long rendered direct, spatially resolved Bell tests on multimode entangled sources experimentally impractical.

\begin{figure}[htbp]
    \centering
     \includegraphics[width=0.8\textwidth]{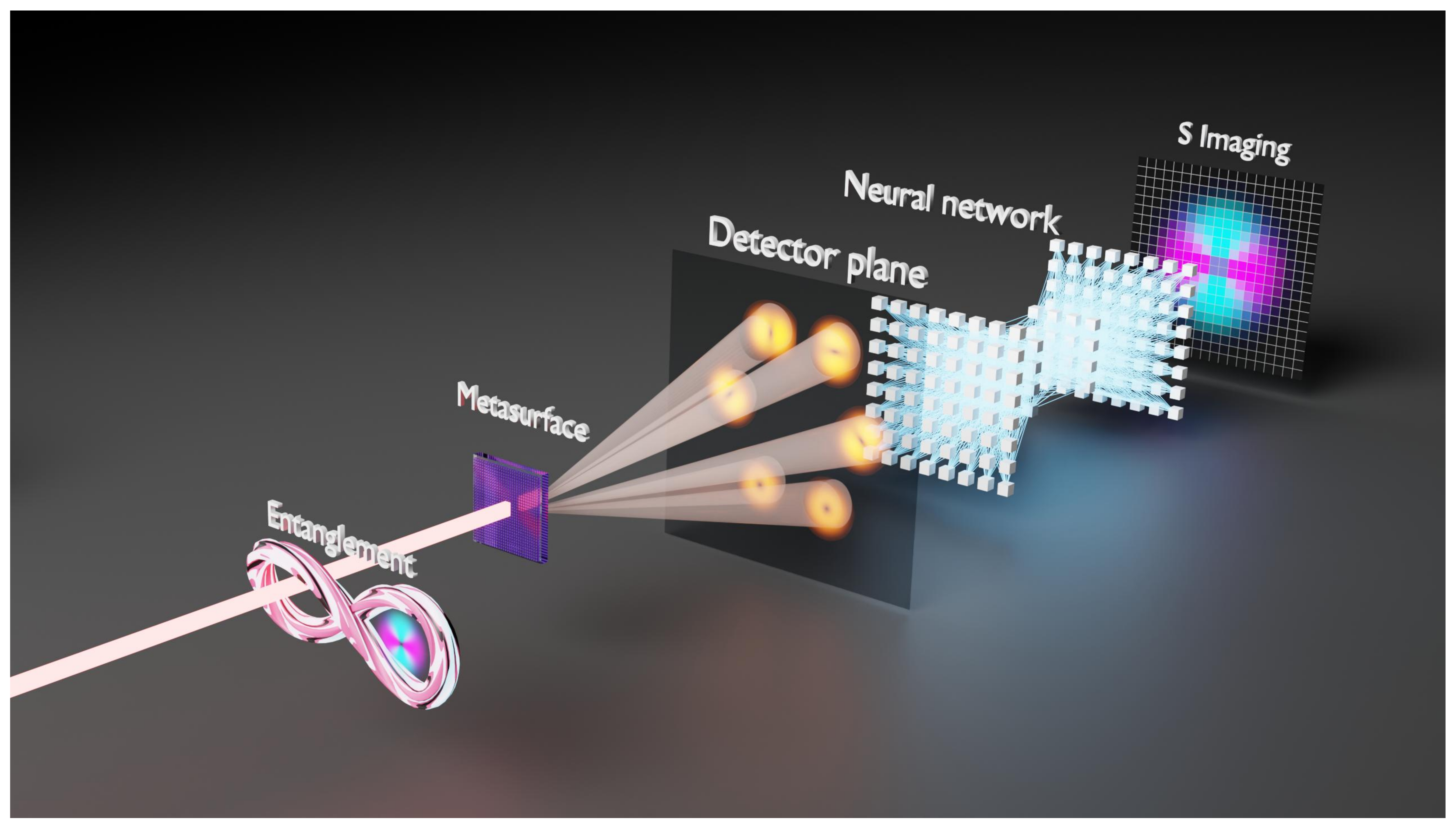}
    \caption{\textbf{Spatially resolved reconstruction of quantum nonlocality using a metasurface and quantum-swin fusion network (Quantum-SFNet).}  A spatially structured entangled photon field is incident on a polarization-analyzing metasurface, which spatially separates Alice's photons according to their Pauli measurement outcomes. The resulting polarization-resolved intensity patterns are projected onto a detection plane and recorded together with photon arrival times and output-port information. From these sparse, photon-limited measurement data, a neural network (Quantum-SFNet) reconstructs the spatial distribution of CHSH nonlocality, producing a pixel-resolved CHSH $S$ map of the entangled field. It enables imaging of spatially distributed nonlocal correlations under ultralow-photon conditions.}
    \label{fig:2}
\end{figure}

To overcome the quadratic-scaling bottleneck, we introduce a hybrid measurement-and-reconstruction framework for spatially resolved reconstruction of quantum nonlocality. Unlike conventional non-imaging Bell tests, our framework is intrinsically based on imaging and operates in an ultralow-photon regime previously challenging for such quantum-correlation measurements. As illustrated in Fig.~\ref{fig:2}, a spatially structured, polarization-entangled biphoton field with position-dependent correlations is incident on a polarization-analyzing metasurface, which performs parallel Pauli projections and maps the measurement outcomes into spatially separated intensity channels on a detection plane. The resulting photon-limited, spatially resolved detection events are then processed by a dedicated neural network architecture named quantum-swin fusion network (Quantum-SFNet), which reconstructs the pixel-wise CHSH parameters from sparse coincidence pairs. In this way, the spatial distribution of Bell nonlocality can be reconstructed and rendered as an image. By combining hardware-level parallelization with physics-informed learning, our scheme realizes spatial nonlocality imaging, with an average of fewer than two detected coincidence pairs per pixel per measurement setting, while preserving both statistical sensitivity and spatial resolution.

\begin{figure}[htbp]
    \centering
     \includegraphics[width=0.9\textwidth]{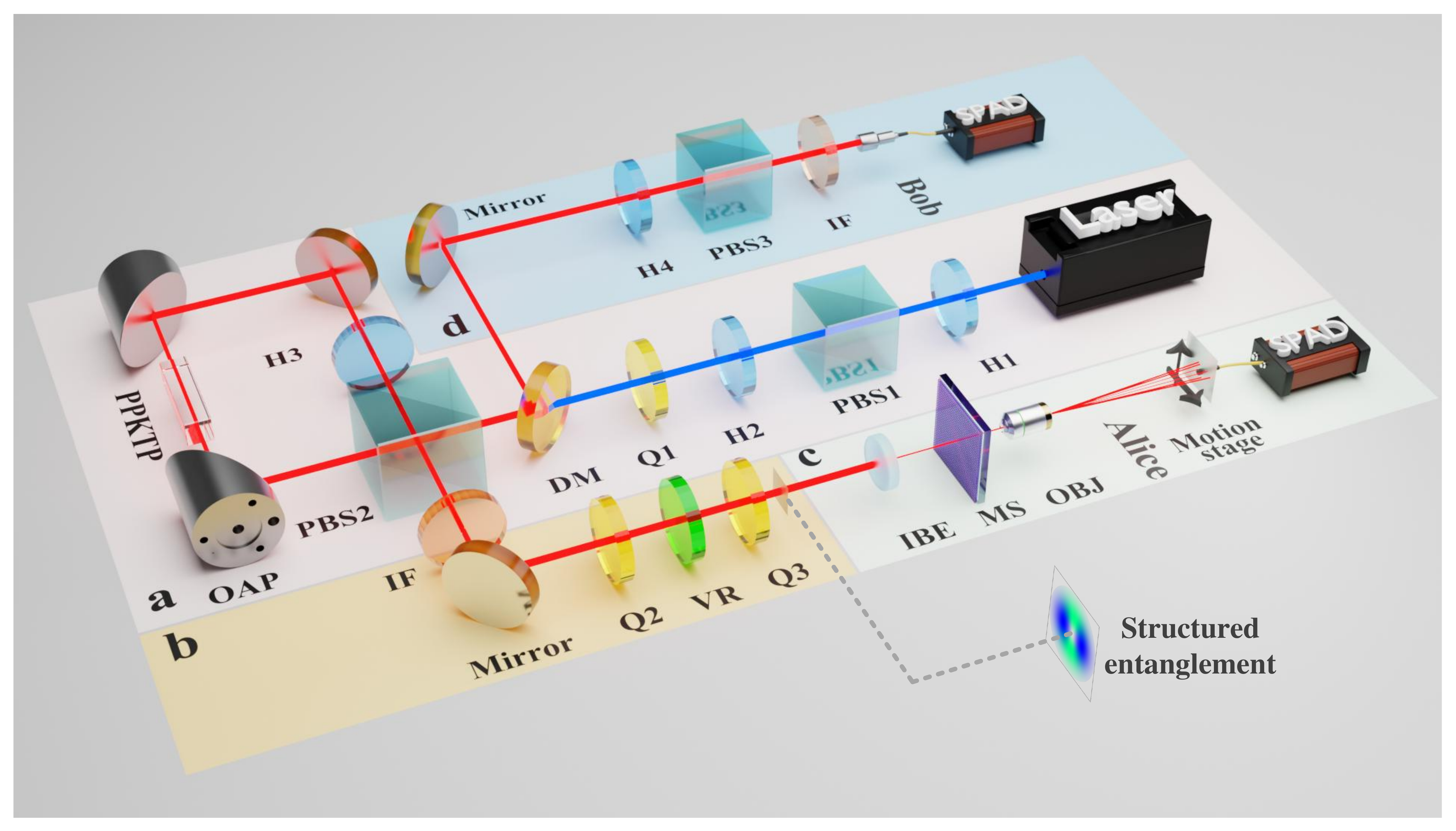}
\caption{\textbf{Experimental setup for spatially resolved nonlocality imaging.}
The setup comprises four modules: source preparation, spatial encoding, and detection (Alice and Bob).
\textbf{a}, Source Preparation (white zone): a 405-nm continuous-wave laser pumps a Sagnac interferometric loop containing a PPKTP crystal, producing polarization-entangled photon pairs $\frac{1}{\sqrt{2}}
\left(
\ket{H_A V_B}
-
\ket{V_A H_B}
\right)$ via type-II spontaneous parametric down-conversion. 
\textbf{b}, Spatial Encoding (yellow zone): the photon directed to Alice undergoes spatial modulation. A sequence comprising quarter-wave plates (Q1--Q3) and a vortex retarder (VR) couple the polarization state to orbital angular momentum (OAM), imprinting a structured entangled field (visualized as the semi-transparent plane labeled as structured entangled field).
\textbf{c}, Alice (spatial detection): The structured photon is expanded by an inverted beam expander (IBE) and projected onto a Pauli-measurement metasurface (MS). An objective lens (OBJ) maps the polarization-resolved diffraction orders onto a detector plane, where a motion stage scans a multimode fiber to record spatially resolved coincidence pairs. 
\textbf{d}, Bob (reference analysis): the partner photon is directed to the reference analysis module (blue zone), and undergoes standard single-mode polarization projection via a half-wave plate (H4) and a polarizing beam splitter (PBS3). 
DM, dichroic mirror; OAP, off-axis parabolic mirror; HWP, half-wave plate; QWP, quarter-wave plate; IF, interference filter; SPAD, single-photon avalanche diode.} 
    \label{fig:3}
\end{figure}

Next, we will provide a detailed description of the experimental realization of this imaging architecture under ultralow coincidence-count conditions. As shown in Fig.~\ref{fig:3}, the experiment comprises four modules: source preparation (Fig.~\ref{fig:3}a), spatial encoding (Fig.~\ref{fig:3}b), and local measurements performed by Alice and Bob (Figs.~\ref{fig:3}c,d). Alice performs spatially resolved Pauli measurements via a metasurface, accessing local observables across multiple spatial modes, while Bob conducts conventional single-mode polarization measurements.

As depicted in Fig.~\ref{fig:3}a, we generate spatially degenerate, polarization-entangled photon pairs via spontaneous parametric down-conversion (SPDC). The 405-nm continuous-wave laser light is injected into a type-II nonlinear crystal, which is pumped in two opposite directions. The relative phase between the two emission paths is stabilized interferometrically, yielding a polarization-maximally-entangled Bell state
\begin{equation}
\frac{1}{\sqrt{2}}(|H_{A}V_{B}\rangle-|V_{A}H_{B}\rangle).
\end{equation}
Details of the entangled source are described in the previous work \cite{mengAIPolarizationCompensation2023, yu2026experimental}. 

To introduce spatial structure while preserving entanglement, the photon sent to Alice passes through a quarter-wave plate--vortex retarder--quarter-wave plate (Q2--VR--Q3) assembly (see Fig.~\ref{fig:3}b), enabling coherent coupling between polarization and OAM. In our experiment, the spatial structure of polarization-entangled state is implemented using Laguerre--Gaussian (LG) modes, expressed as
\begin{equation}
LG_{p,l}\propto\left(\frac{r\sqrt{2}}{w}\right)^{|l|}L_p^{|l|} \left(\frac{2r^2}{w^2}\right)e^{-\frac{r^2}{w^2}}e^{il\theta},
\end{equation}
whose azimuthal phase dependence provides a well-established and widely-used platform for encoding spatial quantum correlations. The prepared state is a coherent superposition of orthogonal polarization components with opposite azimuthal phase profiles:

\begin{equation}\label{eq5}
\rho_{\mathrm{exp}}
=
\ket{\psi_{\mathrm{exp}}}
\bra{\psi_{\mathrm{exp}}},
\ket{\psi_{\mathrm{exp}}}
=
\frac{1}{\sqrt{2}}
\left(
\ket{H_A,\ell_A=+1}\ket{V_B}
-
\ket{V_A,\ell_A=-1}\ket{H_B}
\right).
\end{equation}
Importantly, throughout this work, we consider the experimentally prepared polarization--OAM hybrid entangled state $\rho_{\mathrm{exp}}$ in Eq.~(\ref{eq5}), in which the two polarization components of Alice's photon are coherently coupled to the opposite OAM modes $\ell_A=\pm1$, thereby introducing opposite OAM-dependent spatial phase profiles. When the transverse position $r$ in Alice's detection plane is resolved, the corresponding local two-photon polarization state is denoted by $\rho_{r,\mathrm{exp}}$.

The observable Bell violation is fragile under coarse-grained detection: detection without spatial resolution integrates over the entire field (e.g., via a single-mode fiber) mixes spatial components and suppresses phase-sensitive correlations (i.e., suppresses  observable Bell violations). As the spatial resolution increases, the local structure of $\rho_{r,\mathrm{exp}}$ is progressively recovered, enabling strong and well-resolved nonlocal correlations.

To overcome the suppression of nonlocality induced by coarse-grained detection, we introduce a Pauli-measurement metasurface. By means of spatial multiplexing, this device simultaneously maps the six Pauli projections associated with three mutually orthogonal polarization bases onto six spatially separated focal spots on the detection plane~\cite{renSingleshotFullStokesImaging2025}. In this way, it avoids the sequential basis switching required in conventional time-multiplexed measurements and significantly enhances the efficiency of parallel multi-basis measurements while preserving spatially resolved readout. The experimental configuration is given in Figs.~\ref{fig:3}c and \ref{fig:3}d. In the experiment,  Alice's photon is routed to the metasurface, while its partner undergoes standard polarization analysis. An inverted beam expander adjusts the incident collimated beam to a diameter of $\sim500$~$\mu$m, matching the effective metasurface aperture. The metasurface simultaneously implements six Pauli projection channels, mapping the outcomes onto six spatially separated focal spots arranged concentrically on a common detection plane. This design enables massively parallel, multi-pixel, multi-basis polarization analysis within a single global measurement setting, eliminating the need for sequential basis switching across different spatial modes. An objective lens images the polarization-resolved focal spots onto the detection plane, where Alice collects photons using a motorized translation stage equipped with a multimode fiber (62.5~$\mu$m core diameter), and coincidence measurements are performed jointly with Bob. A complete optical characterization of the metasurface is provided in the Methods and Supplemental Material S2.

\subsection*{Photon-efficient nonlocality reconstruction}
While metasurface-enabled parallelization provides significant measurement convenience by simultaneously mapping multiple bases, overcoming the fundamental photon-efficiency barrier of spatially resolved Bell tests requires further algorithmic innovation. Although recent advances in first-photon and photon-efficient imaging have dramatically expanded the regime in which meaningful information can be extracted from photon-starved measurement data, these approaches focus on reconstructing classical or semiclassical observables (e.g., three-dimensional intensity) and cannot directly access quantum properties.

To overcome this limitation, we develop Quantum-SFNet, a quantum-adaptive inference neural network framework specifically designed to reconstruct the spatially resolved CHSH parameter from starved photonic entanglement measurement data. It integrates massively parallel Pauli measurements enabled by a metasurface with physics-informed learning, processing photon detection events as a unified spatio-temporal data cube under a physics-constrained loss function. This mechanism ensures that the reconstructed correlations remain physically meaningful and interpretable. Importantly, although the ultimate target is the S parameter, Quantum-SFNet does not learn S directly. It first reconstructs intensity images for the 16 projection configurations under physical constraints and then computes S from those images. The loss function enforces these physical constraints by separately optimizing the real and imaginary parts of off-diagonal terms to preserve Hermiticity, while positive semidefiniteness is promoted through trace-distance-based regularization. In the following analysis, the experimentally generated state remains the polarization--OAM hybrid entangled state given in Eq.~(\ref{eq5}). The target state $\rho_t$ used to construct the CHSH measurement operator is chosen as
\begin{equation}\label{eq6}
\rho_t(\varphi)
=
\ket{\psi_t(\varphi)}
\bra{\psi_t(\varphi)},
\end{equation}
with
\begin{equation}
\ket{\psi_t(\varphi)}
=
\frac{1}{\sqrt{2}}
\left(
\ket{H_A V_B}
+
e^{i\varphi}
\ket{V_A H_B}
\right),
\end{equation}
where $\varphi$ determines the relative phase of the reference Bell state used for the CHSH measurement construction.

\begin{figure}[htbp]
    \centering
     \includegraphics[width=0.8\textwidth]{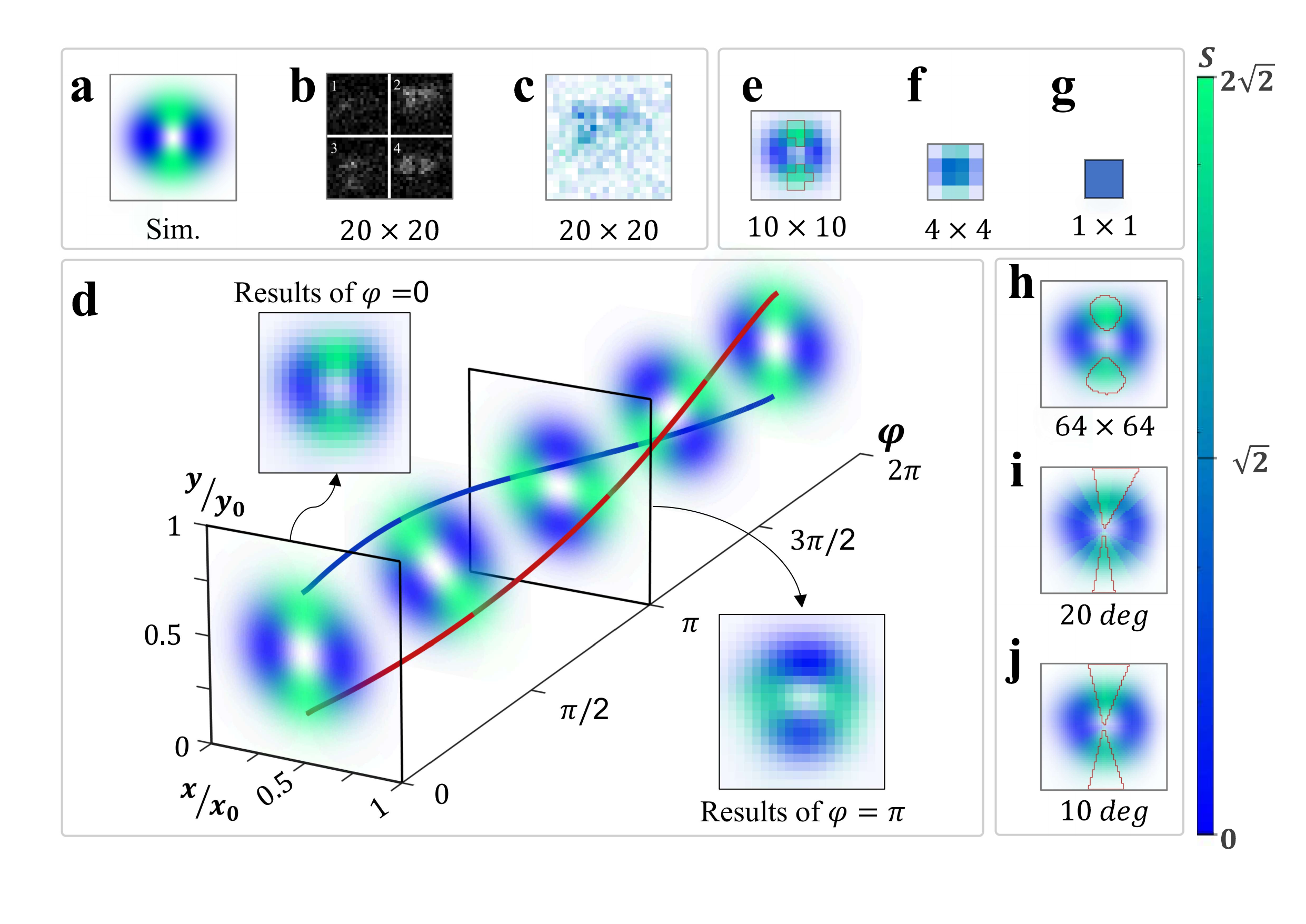}
   \caption{\textbf{Experimental reconstruction and image-based analysis of spatially distributed Bell nonlocality.} 
\textbf{a}, theoretical CHSH parameter ($S$-value) distribution of the prepared spatially structured polarization--OAM hybrid entangled states.
\textbf{b}, raw photon-count data (intensity maps) recorded under ultralow-photon conditions (average of 1.70 coincidence pairs per pixel per projection configuration); no spatial structure is discernible. 
\textbf{c}, direct CHSH parameter reconstruction from raw data without neural-network enhancement; severe statistical fluctuations completely obscure the underlying nonlocality.
\textbf{d},  Simulated and experimentally reconstructed spatial CHSH distributions for different target-state phases $\varphi$ in Eq.~(\ref{eq6}). As $\varphi$ varies, the location of the maximum Bell violation rotates accordingly (see the blue and red trajectories), demonstrating the phase-dependent nature of nonlocality imaging. The insets show the reconstructed results for $\varphi=0$ and $\varphi=\pi$. The regions enclosed with red outlines exhibit genuine Bell violations ($S>2$). 
\textbf{e--g}, the reconstructed coarse-grained CHSH maps at resolution of $10\times10$, $4\times4$, and $1\times1$ (emulating progressively reduced spatial resolution), in which the violations gradually disappear.
\textbf{h--j}, Spatial upsampling and angular structural analysis: the reconstructed $20\times20$ map is first upsampled to \textbf{h} $64\times64$ pixels for continuous geometry visualization, and angular integration is then applied by calculating the effective $S$ value within discrete azimuthal sectors of \textbf{i} $20^\circ$ and \textbf{j} $10^\circ$, confirming the rotational symmetry and the angularly localized Bell violations.}
    \label{fig:4}
\end{figure}

Figure~\ref{fig:4} presents the theoretical spatial nonlocality distribution, together with the corresponding measurement data and nonlocality imaging results, for the polarization--OAM hybrid entangled state generated in Eq.~(\ref{eq5}). In both the simulations and the experimental data processing, except for Fig.~\ref{fig:4}d, the CHSH measurement operators are constructed using the reference target state with relative phase $\varphi=0$.
 The theoretical nonlocality spatial distribution of the input polarization--OAM hybrid entangled state is shown in Fig.~\ref{fig:4}a. Here, 16 projection configurations are used. Figure~\ref{fig:4}b displays representative measurement data collected from different output ports of the metasurface when Bob measures linear polarization at $22.5^\circ$ (denoted as $\Pi_B=22.5^\circ$); this setting is chosen solely for illustration, while other polarization (including circular) bases are employed in different measurement tasks discussed below. This photon-efficient algorithm achieves efficient nonlocality reconstruction of a 400-pixel entangled field within a single experimental run. It allows direct observation of the resolution-induced degradation of nonlocality, in quantitative agreement with Eq.~(\ref{eq2}). Across all measurements, a maximum of 8 photons are detected per pixel over 64 time bins, corresponding to 1.70 coincidence pairs per pixel per projection configuration, circumventing the thousands of detection events per spatial mode typically required in conventional Bell tests. At this ultralow photon flux, the raw two-dimensional intensity maps fail to exhibit any discernible orbital angular momentum (OAM)-related spatial features (see Fig.~\ref{fig:4}c). After processing the data with Quantum-SFNet, the reconstructed intensity images for all 16 projection configurations yield a spatial $S$-value distribution. Figure~\ref{fig:4}d shows the simulated and neural-network-processed experimental results for CHSH measurements performed on the polarization--OAM hybrid entangled state in Eq.~(\ref{eq5}), using CHSH measurement operators constructed from target states with different relative phases $\varphi$. As $\varphi$ varies, the position of the maximum $S$ value rotates accordingly. In particular, for $\varphi=0$ and $\varphi=\pi$, the strongest Bell violations appear predominantly along the vertical and horizontal directions, respectively, consistent with the simulation predictions and directly demonstrating the target-state-dependent spatial distribution of nonlocality. In the network-processed result, the red-contoured region exhibits clear violations ($S>2$), with a maximum observed nonlocality of $S_{\max}=2.66$.

\subsection*{Image-based structural analysis and versatile applications}

Having access to the reconstructed nonlocality image, we can analyze Bell nonlocality directly in an image-based manner. To study how spatial resolution governs observable Bell nonlocality, an effect inaccessible to conventional single-channel measurements, we apply controlled spatial downsampling to the reconstructed CHSH maps to simulate detector resolution with increasing effective pixel sizes (see Figs.~\ref{fig:4}e--\ref{fig:4}g). As the resolution decreases, the contrast of the nonlocality distribution degrades and the maximum CHSH $S$ value drops monotonically. At native $20\times20$ resolution, a strong violation is observed ($S_{\max}=2.66$); at $10\times10$ and $4\times4$ resolution, the violation becomes marginal; and at $1\times1$ single-pixel limit, no observable Bell violation remains ($S_{\max}=1.18$ falls below the classical bound). This highlights an important physical insight: in spatially structured entangled fields, coarse graining leads to incoherent spatial averaging of distinct correlation regions, and may thereby conceal genuine nonlocality. Spatial resolution is therefore not merely a technical refinement, but a fundamental requirement for revealing quantum nonlocality distribution. Thus, our framework preserves the spatial structure of distributed nonlocal correlations rather than reducing them to a single global Bell parameter.
Representing nonlocal correlations in an image form also enables richer structural analysis of the underlying entangled field. The LG-entangled photon pairs exhibit intrinsic rotational symmetry: for a fixed azimuthal angle $\theta$, the local polarization state $\rho_{r,\mathrm{exp}}$ is independent of the radial coordinate, up to normalization. To exploit this property, we first upsample the reconstructed nonlocality map to obtain sufficient spatial resolution (see Fig.~\ref{fig:4}h), and then perform angular downsampling by integrating pixel values along different azimuthal directions. The resulting angular dependence of the CHSH parameter $S(\theta)$, as shown in Figs.~\ref{fig:4}i and \ref{fig:4}j, reveals that the violation of local realism is concentrated around $\theta = \pi/2, 3\pi/2$, in excellent agreement with theoretical expectations. At large radii, however, the photon flux becomes too weak to yield reliable statistics, leading to a reduced angularly averaged nonlocality of $S_{\max}=2.29$. This image-based analysis highlights how a spatially structured entangled field can encode directional signatures of nonlocality that would remain hidden in single-channel measurements.

\begin{figure}
    \centering
    \includegraphics[width=1\linewidth]{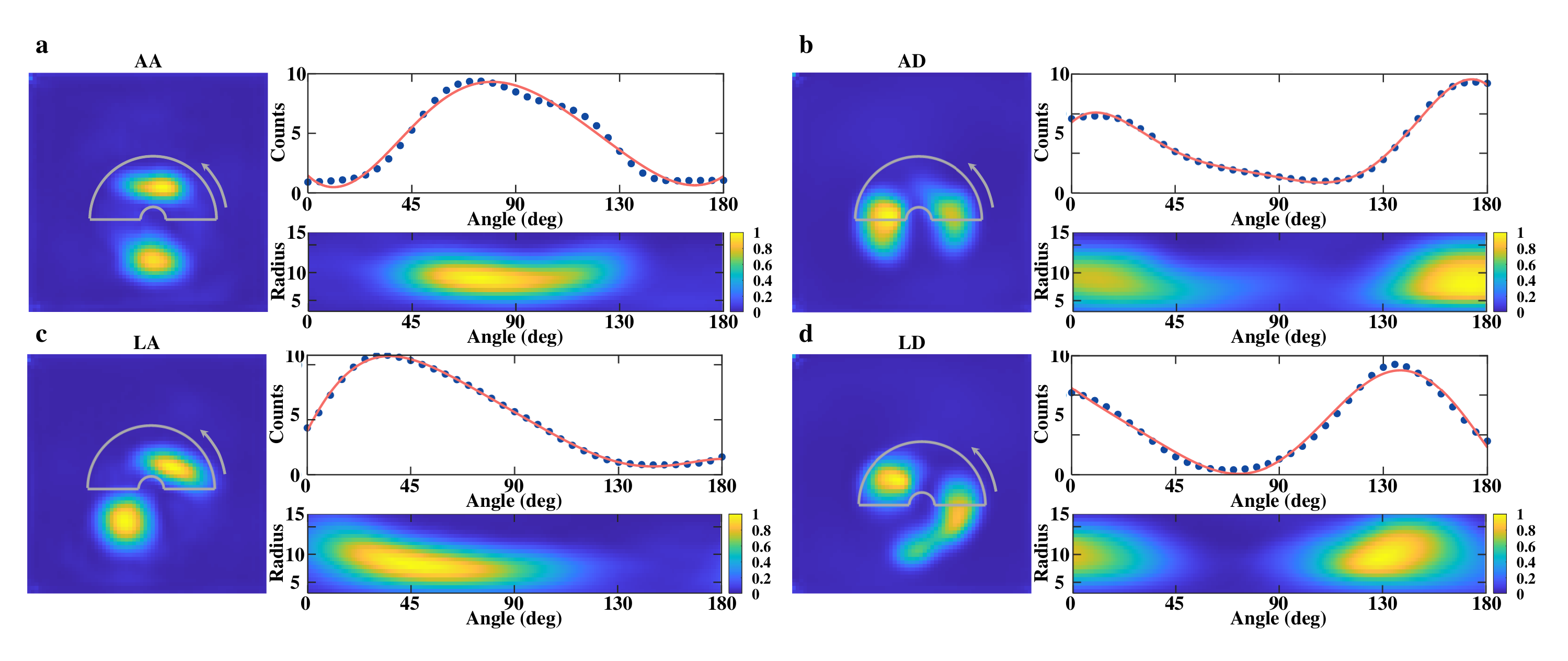}
    \caption{\textbf{Reconstruction of Bell-type nonlocality via Quantum-SFNet-assisted imaging}. \textbf{a}–\textbf{d}, Experimental results under four distinct two-photon polarization bases ($\Pi_{A}=A,L;\Pi_{B}=A,D$). The left panels display the normalized 2D spatial intensity maps output by the neural network. The red annular boundaries define the region of interest (ROI), with the solid red line marking the azimuthal zero-angle reference ($\theta = 0^\circ$). The bottom-right panels show the corresponding unwrapped polar-to-Cartesian intensity distributions (Radius vs. $\theta$). The top-right panels plot the integrated angular coincidence counts (blue dots) along with their sinusoidal fits (red curves). Using the correlated intensities at specific optimal angles derived from these curves, the Bell inequality violation is verified with a calculated CHSH parameter of $S=2.52$.}
    \label{fig:5}
\end{figure}

This experimental architecture is also versatile and can be extended to other quantum information protocols by appropriately selecting output ports and local measurement settings. Global Bell-type nonlocal behavior can be extracted from full-field quantum imaging data ~\cite{moreauImagingBelltypeNonlocal2019}. Once the output measurement images are obtained, our framework can be naturally connected to this global analysis method. Although our framework is primarily designed for spatially resolved nonlocality reconstruction, the reconstructed measurement images can also be incorporated into existing global Bell-type imaging analyses. As shown in Fig.~\ref{fig:5}, different choices of the polarization channels produce intensity patterns analogous to those reported in the previous work. From these azimuthally resolved intensity distributions, one can reconstruct a global CHSH parameter $S$ using a previously established global imaging analysis procedure. In our setting, this parameter probes Bell-type correlations in an effective hybrid two-qubit space, namely between the biphoton polarization logical qubit encoded in $\mathrm{span}\{|H_{A}V_{B}\rangle,|V_{A}H_{B}\rangle\}$ and the OAM qubit encoded in $\mathrm{span}\{|\ell_A=+1\rangle,|\ell_A=-1\rangle\}$. Using this global analysis, we obtain $S=2.52>2$. This result not only further confirms the quality of the entangled state generated in our scheme, but also shows that the proposed nonlocal imaging framework can access both pixel-resolved and global manifestations of Bell nonlocality across spatial modes when combined with existing global Bell-type nonlocality imaging analyses.

As another example of the versatility of our framework, we further demonstrate remote state preparation (RSP) using the same spatially structured entangled photon source. By fixing Bob's polarization projection $\Pi_B$ to a predefined state and recording coincidence events at Alice's six metasurface output ports, spatially distributed single-photon states are conditionally prepared and characterized on Alice's side. Specifically, projecting Bob's photon onto the $\Pi_B=A$ or $L$ polarization state remotely prepares the corresponding spatially distributed state of Alice's photon. Full quantum state tomography is then performed on Alice's photon using the spatially resolved polarization-projection results recorded through the metasurface, without any modification of the optical hardware. This demonstrates the generality of our approach and indicates that spatially resolved nonlocality measurements and quantum-state engineering can be unified within a single photon-efficient platform. Further details of the neural-network architecture, ablation studies, additional experimental results, and analyses are provided in Supplementary Materials~S3 and S4.

\section*{Discussion and conclusion}

In this work, we establish a new experimental framework for nonlocality imaging in which spatially resolved Bell measurements become image based, parallelized, and feasible in the ultralow photon regime. We first investigate the influence of spatial resolution on nonlocality imaging. Higher spatial resolution enables distributed nonlocal correlations to be more faithfully resolved and avoids the loss of observable Bell violations caused by coarse-grained detection. However, increasing the resolution also leads to a severe growth in resource consumption, which has long made it difficult to observe Bell nonlocality as a spatially distributed observable. To address this challenge, our framework combines a metasurface-enabled Pauli measurement with photon-efficient imaging, enabling spatially resolved Bell tests with substantially reduced measurement settings and photon counts, and successfully realizing nonlocality imaging across an entangled field. Remarkably, each pixel requires only $\sim27.2$ (i.e., $1.7\times16$) detected coincidence pairs, and the full 400-pixel map needs merely $\sim1.1\times10^{4}$ (i.e., $1.7\times16\times400$) detected coincidence pairs. In contrast, conventional single-mode approaches cannot directly preserve the spatial structure of multimode Bell correlations, since spatial integration across multiple modes washes out observable Bell violations.  Building on this framework, we further investigate the target-state dependence of CHSH measurements. Finally, we demonstrate the compatibility of the framework with existing quantum imaging schemes, as well as its versatility for multiple applications.

In comparison with recent high-performance experiments~\cite{cabelloLoopholeFreeBellTest2012, christensenDetectionLoopholeFreeTestQuantum2013, giustinaSignificantLoopholeFreeTestBells2015, larssonLoopholesBellInequality2014, yinSatellitebasedEntanglementDistribution2017a,moreauImagingBelltypeNonlocal2019,shalmStrongLoopholeFreeTest2015a}, our method pushes Bell verification into an extremely photon-starved regime, requiring only a minimal number of detected coincidence pairs per pixel to reveal nonlocality. Remarkably, with a total photon budget of only $\sim 10^4$, which is already comparable to the count level used in many single-mode nonlocality measurements, we reconstruct a 400-pixel CHSH image of an entangled field, realizing image-based characterization of spatially distributed Bell nonlocality. By preserving spatial resolution, our imaging framework is suited for spatially distributed entanglement in quantum computation, networking, and large-scale photonic technologies.

This scalable, resource‑efficient method provides a new pathway for visual characterization and utilization of complex entangled states. Beyond Bell nonlocality, it can extend to other quantum tasks such as remote state preparation and image‑based quantum state analysis, highlighting its versatility as a general‑purpose quantum imaging tool. The approach can be further integrated with emerging platforms (e.g., two-dimensional material sources, large‑area photonic arrays) to characterize multi‑pixel emitters. Representing quantum correlations as images also offers an intuitive and powerful language for understanding and diagnosing high‑dimensional quantum systems, and we believe that spatial nonlocality imaging will play an important role in bridging abstract quantum resources and their practical implementations in large‑scale quantum networks.

\section*{Methods}
\subsection*{Metasurface design and spatial multiplexing}

We employ a Pauli-measurement-multiplexed metasurface to enable highly parallel Pauli measurements. By harnessing spatial multiplexing, an aperiodic array of anisotropic silicon nanopillars enables parallel polarization analysis across three mutually orthogonal polarization bases. As a result, the six Pauli eigenstates ($\ket{H}$, $\ket{V}$, $\ket{L}$, $\ket{R}$, $\ket{D}$, $\ket{A}$) are deterministically encoded into six spatially resolved focal spots arranged concentrically on a common focal plane. The device is engineered with an effective aperture area exceeding $0.25~\mathrm{mm}^2$ to ensure full coverage of the incident structured beam. Accurate spatial separation among the six focal spots is achieved by prescribing sufficiently large inter-spot spacings, thereby facilitating subsequent fiber-coupled single-photon detection. Detailed design principles, the Jones-matrix formalism, phase-profile equations, and finite-difference time-domain (FDTD) simulation parameters are provided in Supplementary Materials~S2.

\subsection*{Metasurface fabrication and optical performance characterization}

The metasurface was fabricated on an epitaxial silicon-on-sapphire platform using electron-beam lithography followed by silicon dry etching. Its optical performance was evaluated by measuring the polarization-state recognition accuracy and the Stokes-parameter reconstruction errors of each channel. Experimental characterization confirms that the metasurface concurrently achieves highly parallelized polarization demultiplexing and full-aperture coverage. Full fabrication details, morphological characterization, and classical optical characterization of the six-channel analyzer---including focal-plane measurements under six canonical polarization inputs, Bloch-sphere mapping, and vector-vortex-beam validation---are provided in Supplementary Materials~S2 and Fig.~S1.

\section*{Acknowledgements}
This work was supported by the National Natural Science Foundation of China (Grant Nos.~ 12474480, 92365115, and 62571047), the Beijing Natural Science Foundation (Grant Nos.~1262037, F261018). We thank Dr Xian-Song Ren for his valuable help in the design of the metasurface.

\section*{Author contributions}
J.L. and Prof.~A.-N.Z. designed the scheme; Prof.~X.G., Z.-M.F. fabricated the metasurface; Q.-Y.W. and J.L. developed the neural network; J.L., Z.-M.F., Q.-Y.W., Prof.~W.-K.Y., Z.M., X.-Y.F., W.-H.W., and J.M. performed the experiment. All of the authors contributed to the writing of the manuscript.

\section*{Additional information}
The authors declare no competing interests. Supplementary information is available for this paper. 

Correspondence should be addressed to A.-N.Z. .

\bibliography{reference.bib} 

\end{document}


\maketitle

\section{Target-State-Dependent CHSH Measurement Settings}

For the CHSH inequality measurements discussed in this work, the Bell parameter is written as
\begin{equation}
S
=
E(a,b)+E(a',b)+E(a,b')-E(a',b'),
\end{equation}
where $a$ and $a'$ denote the two local measurement settings on Alice's polarization qubit, while $b$ and $b'$ denote those on Bob's polarization qubit. The correlation function is defined as
\begin{equation}
E(a,b)
=
\frac{
C(a,b)+C(a^{\perp},b^{\perp})
-C(a^{\perp},b)-C(a,b^{\perp})
}{
C(a,b)+C(a^{\perp},b^{\perp})
+C(a^{\perp},b)+C(a,b^{\perp})
},
\end{equation}
where $C$ denotes the measured coincidence counts under the corresponding joint projective measurement settings.

In our spatially resolved nonlocality imaging framework, Alice's photon is analyzed using the polarization-projection metasurface, whereas Bob's photon undergoes conventional single-mode polarization analysis. Accordingly, Alice's two measurement settings are fixed as
\begin{equation}
a=\ket{H},
\qquad
a'=\ket{D},
\end{equation}
where
\begin{equation}
\ket{D}
=
\frac{1}{\sqrt{2}}
\left(
\ket{H}+\ket{V}
\right).
\end{equation}

For the target Bell state
\begin{equation}
\ket{\psi_t(0)}
=
\frac{1}{\sqrt{2}}
\left(
\ket{H_A V_B}
+
\ket{V_A H_B}
\right),
\end{equation}
the corresponding optimal measurement settings on Bob's photon can be chosen as
\begin{equation}
\begin{aligned}
b
&=
\cos(67.5^\circ)\ket{H}
+
\sin(67.5^\circ)\ket{V},\\
b'
&=
\cos(112.5^\circ)\ket{H}
+
\sin(112.5^\circ)\ket{V}.
\end{aligned}
\end{equation}

More generally, the target state considered in the spatially resolved CHSH measurements is
\begin{equation}
\ket{\psi_t(\varphi)}
=
\frac{1}{\sqrt{2}}
\left(
\ket{H_A V_B}
+
e^{i\varphi}\ket{V_A H_B}
\right).
\end{equation}
The relative phase factor can equivalently be incorporated into a local phase rotation on Bob's polarization qubit. Because Alice's measurement settings are fixed by the metasurface architecture, the target-state dependence is implemented by rotating Bob's reference measurement bases. The corresponding measurement settings are
\begin{equation}
\begin{aligned}
b(\varphi)
&=
\cos(67.5^\circ)\ket{H}
+
e^{-i\varphi}\sin(67.5^\circ)\ket{V},\\
b'(\varphi)
&=
\cos(112.5^\circ)\ket{H}
+
e^{-i\varphi}\sin(112.5^\circ)\ket{V}.
\end{aligned}
\end{equation}

When $\varphi=0$, these settings reduce to the linear-polarization bases given above. As $\varphi$ varies, Bob's reference measurement bases rotate accordingly, giving rise to the target-state-dependent spatial nonlocality distributions reported in the main text.

\section{Parallel Pauli Projection Metasurface Design and Characterization}

To enable parallel Pauli projections within a single planar optical element, we employ a polarization-multiplexed metasurface that simultaneously maps three pairs of mutually orthogonal polarization bases, $(|H\rangle, |V\rangle)$, $(|L\rangle, |R\rangle)$ and $(|D\rangle, |A\rangle)$, into six
spatially separated focal spots on the focal plane. By prescribing polarization-dependent target phase profiles across the metasurface plane, different input polarization states are focused onto predefined spatial positions, thereby encoding polarization projection outcomes into spatially distinguishable output channels. As a result, the incident beam is deterministically routed to six polarization-resolved focal spots, each corresponding to a canonical Pauli eigenstate, thereby providing the hardware foundation for subsequent parallel local measurements.

The metasurface adopts a spatial multiplexing strategy in which each unit cell consists of three birefringent silicon nanopillars, with the three constituent nanopillars assigned to different orthogonal polarization basis pairs. The realization of the target phase profile relies on the coordinated design of the geometry, orientation, and arrangement of these birefringent silicon nanopillars. The optical response of an individual birefringent silicon nanopillar can be described by the Jones matrix
\begin{equation*}
R(\theta)
\begin{pmatrix}
t_u e^{i\varphi_u} & 0 \\
0 & t_v e^{i\varphi_v}
\end{pmatrix}
R(-\theta),
\end{equation*}
where $t_u$ and $t_v$ denote the amplitudes of the co-polarized responses along the fast
and slow axes, respectively, $\varphi_u$ and $\varphi_v$ represent the corresponding phase delays,
and $\theta$ is the in-plane rotation angle of the nanopillar. For orthogonal linear
polarization basis pairs, independent phase control is primarily achieved by aligning
the fast and slow axes with the corresponding polarization bases through $\theta$. For left-
and right-handed circular polarization multiplexing, nanopillars exhibiting quarter-
wave-plate functionality are selected, and independent demultiplexing is realized
through the combined use of the propagation phase, determined by the nanopillar
geometry, and the geometric phase, governed by $\theta$. In this way, a single-layer
metasurface enables six-channel parallel polarization analysis on a common focal plane,
as illustrated in Supplementary Fig.~\ref{fig:meta}a.

To implement six-channel parallel polarization analysis, the target phase profiles are globally engineered according to Eqs.~\ref{var1} and \ref{var2}, where $\varphi_1(x,y)$ corresponds to the phase profiles associated with the $|H\rangle$, $|L\rangle$, and $|D\rangle$ polarization states, while $\varphi_2(x,y)$ corresponds to those associated with the $|V\rangle$, $|R\rangle$, and $|A\rangle$ polarization states.

\begin{equation}\label{var1}
\varphi_1(x,y)=
\begin{cases}
-\dfrac{2\pi}{\lambda_0}\left(\sqrt{(x-a_0)^2+(y-\sqrt{3}a_0)^2+f^2}-f\right), & x=P(3m+1),\ y=Pm, \\
-\dfrac{2\pi}{\lambda_0}\left(\sqrt{(x-2a_0)^2+y^2+f^2}-f\right), & x=P(3m+2),\ y=Pm, \\
-\dfrac{2\pi}{\lambda_0}\left(\sqrt{(x-a_0)^2+(y+\sqrt{3}a_0)^2+f^2}-f\right), & x=P(3m+3),\ y=Pm.
\end{cases}
\end{equation}

\begin{equation}\label{var2}
\varphi_2(x,y)=
\begin{cases}
-\frac{2\pi}{\lambda_0}\left(\sqrt{(x+a_0)^2+(y-\sqrt{3}a_0)^2+f^2}-f\right), & x=P(3m+1),\ y=Pm, \\
-\frac{2\pi}{\lambda_0}\left(\sqrt{(x+2a_0)^2+y^2+f^2}-f\right), & x=P(3m+2),\ y=Pm, \\
-\frac{2\pi}{\lambda_0}\left(\sqrt{(x+a_0)^2+(y+\sqrt{3}a_0)^2+f^2}-f\right), & x=P(3m+3),\ y=Pm.
\end{cases}
\end{equation}

Here, $\lambda_0 = 808\,\mathrm{nm}$ is the operating wavelength, and $f = 1.7\,\mathrm{mm}$ is the focal length. The variables $x$ and $y$ denote spatial positions on the metasurface plane. The lattice period of the nanopillar unit cell is $P = 390\,\mathrm{nm}$, the nanopillar rows are indexed by $m = 0,1,2,\ldots$, and $a_0 = 34\,\mu\mathrm{m}$ denotes the focal-spot offset on the focal plane. By appropriately selecting the target focal-spot positions for different polarization channels, the three orthogonal polarization bases can be stably mapped onto six spatially separated focal spots, thereby enabling parallel Pauli projections.

To establish the optical response database required for device design, systematic finite-difference time-domain (FDTD) parameter sweeps were carried out. In the simulations, the nanopillar height was fixed at 600 nm, the lattice period was set to $390 \mathrm{nm}$, and the nanopillar width was varied from 90 to $240 \mathrm{nm}$. Furthermore, based on a figure-of-merit (FOM) matching algorithm, the optimal nanopillar dimensions and rotation angle at each lattice site were selected, thereby ensuring that the overall phase profile of the device closely matches the target design.

The metasurface was fabricated by electron-beam lithography followed by silicon dry etching. Specifically, a $174\,\mathrm{nm}$-thick layer of PMMA 950K-A4 was spin-coated onto a substrate consisting of a $600\,\mathrm{nm}$-thick epitaxial single-crystalline silicon layer grown on a $475\,\mu\mathrm{m}$-thick sapphire wafer. The sample was prebaked at $180\,^{\circ}\mathrm{C}$ for $10\,\mathrm{min}$ to remove residual solvent and enhance adhesion between the resist and the substrate. The pattern was then directly written by electron-beam lithography with an exposure dose of $740\,\mu\mathrm{C}/\mathrm{cm}^2$, followed by development in MIBK:IPA $=1:3$ for $50\,\mathrm{s}$ and rinsing in IPA for $30\,\mathrm{s}$ to complete pattern transfer. The electron-beam resist was subsequently removed by ultrasonication in a heated acetone bath for $24\,\mathrm{h}$. A $30\,\mathrm{nm}$-thick alumina layer was then deposited by ion-beam sputtering to serve as a hard mask. Finally, deep silicon etching was performed for $220\,\mathrm{s}$ to transfer the pattern into the $600\,\mathrm{nm}$-thick single-crystalline silicon layer. The morphology of the fabricated device was characterized by optical microscopy and scanning electron microscopy (SEM), as shown in Supplementary Fig.~\ref{fig:meta}b, confirming the high fabrication fidelity of the realized structures. The device has a side length of $540\,\mu\mathrm{m}$, which is sufficient to provide full-aperture coverage of the incident beam.

To verify the six-channel polarization analysis capability of the device, we measured the focal-plane output patterns under illumination with six fundamental polarization states, namely $|H\rangle$, $|V\rangle$, $|L\rangle$, $|R\rangle$, and $|D\rangle$, $|A\rangle$, and reconstructed Stokes parameters, as shown in Supplementary Fig.\ref{fig:meta}c. The experimentally measured polarization states agree well with the designed target states, with average absolute errors of $MAE(S_1)=3.29\%$, $MAE(S_2)=2.33\%$, and $MAE(S_3)=1.48\%$, demonstrating the high accuracy of the device for six-channel polarization analysis. In addition, the capability of the device to analyze structured polarization fields was validated using first- and second-order vector vortex beams, as shown in Supplementary Figs.~\ref{fig:meta}d,e. From left to right, the subpanels show the spatially resolved focal-spot distributions, the total intensity distribution, and the reconstructed Pauli parameter maps obtained from the six-channel intensities, namely $\sigma_z=(I_H-I_V)/(I_H+I_V)$, $\sigma_x=(I_D-I_A)/(I_D+I_A)$, and $\sigma_y=(I_L-I_R)/(I_L+I_R)$. These results demonstrate that the six-focal-spot metasurface enables highly parallelized Pauli projections under full-aperture illumination.

\begin{figure}[htbp]
\centering

\includegraphics[width=0.8\textwidth]{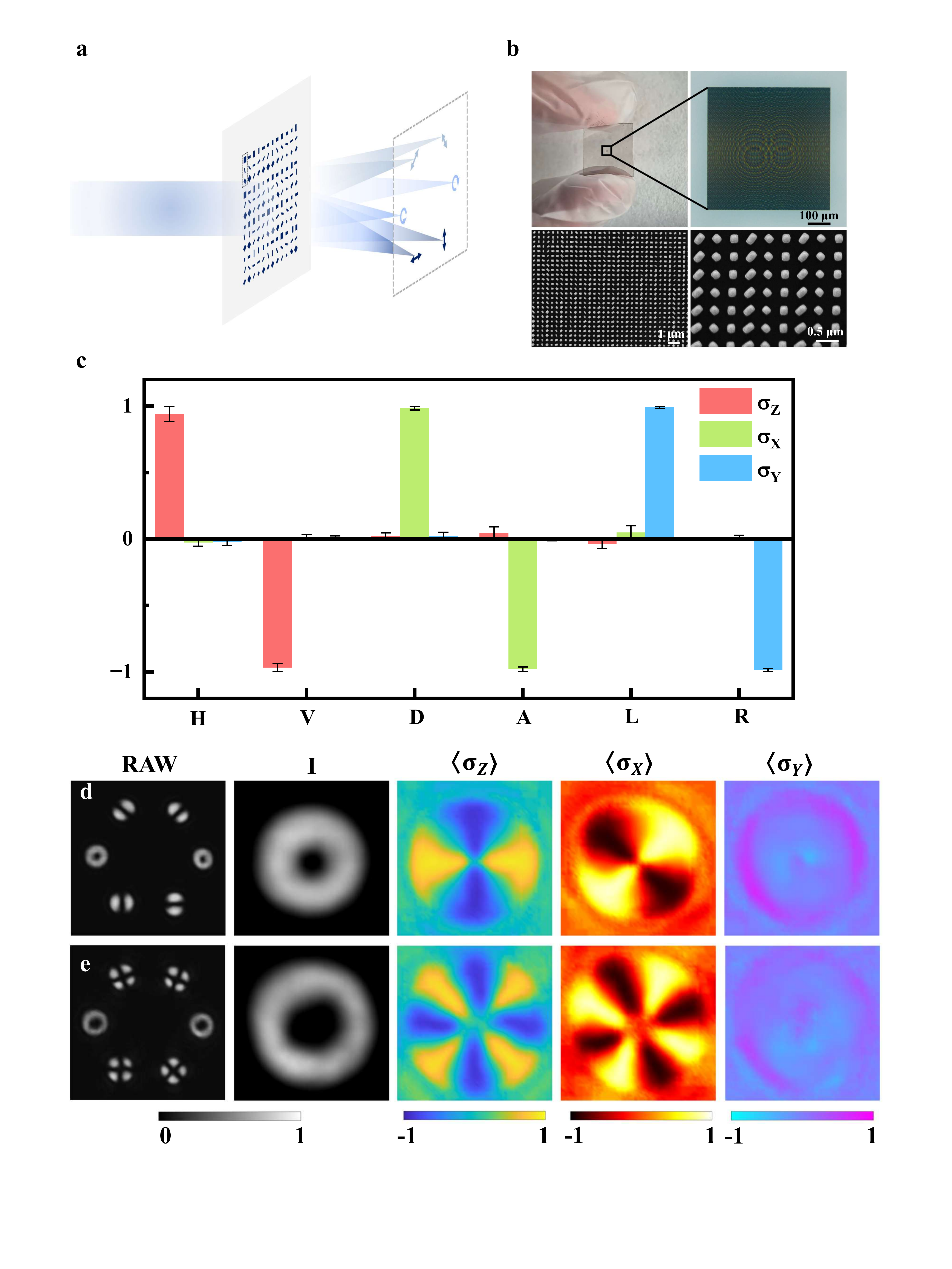}
\caption{\textbf{Design, fabrication, and optical characterization of the polarization-multiplexed metasurface}. \textbf{a} Schematic illustration of the spatial multiplexing principle. The incident beam is deterministically routed into six spatially resolved focal spots, each corresponding to a canonical Pauli eigenstate. \textbf{b} Photograph of the fabricated metasurface, optical microscope image, and scanning electron microscope (SEM) image, showing the top-view morphology of the silicon nanopillars. \textbf{c} Full Stokes parameters under different incident polarizations, with error bars representing the deviation between calculated and theoretical values. \textbf{d,e} Validation of the polarization imaging capability of the device using first-order \textbf{d} and second-order \textbf{e} vector vortex beams, respectively. From left to right, the panels show the spatially resolved focal spots, the total intensity, and the corresponding reconstructed Pauli parameter maps $\langle \sigma_z \rangle$, $\langle \sigma_x \rangle$, and $\langle \sigma_y \rangle$.}

\label{fig:meta}
\end{figure}

\section{Quantum-SFNet Architecture}

\begin{figure}[htbp] \centering \includegraphics[width=0.95\textwidth]{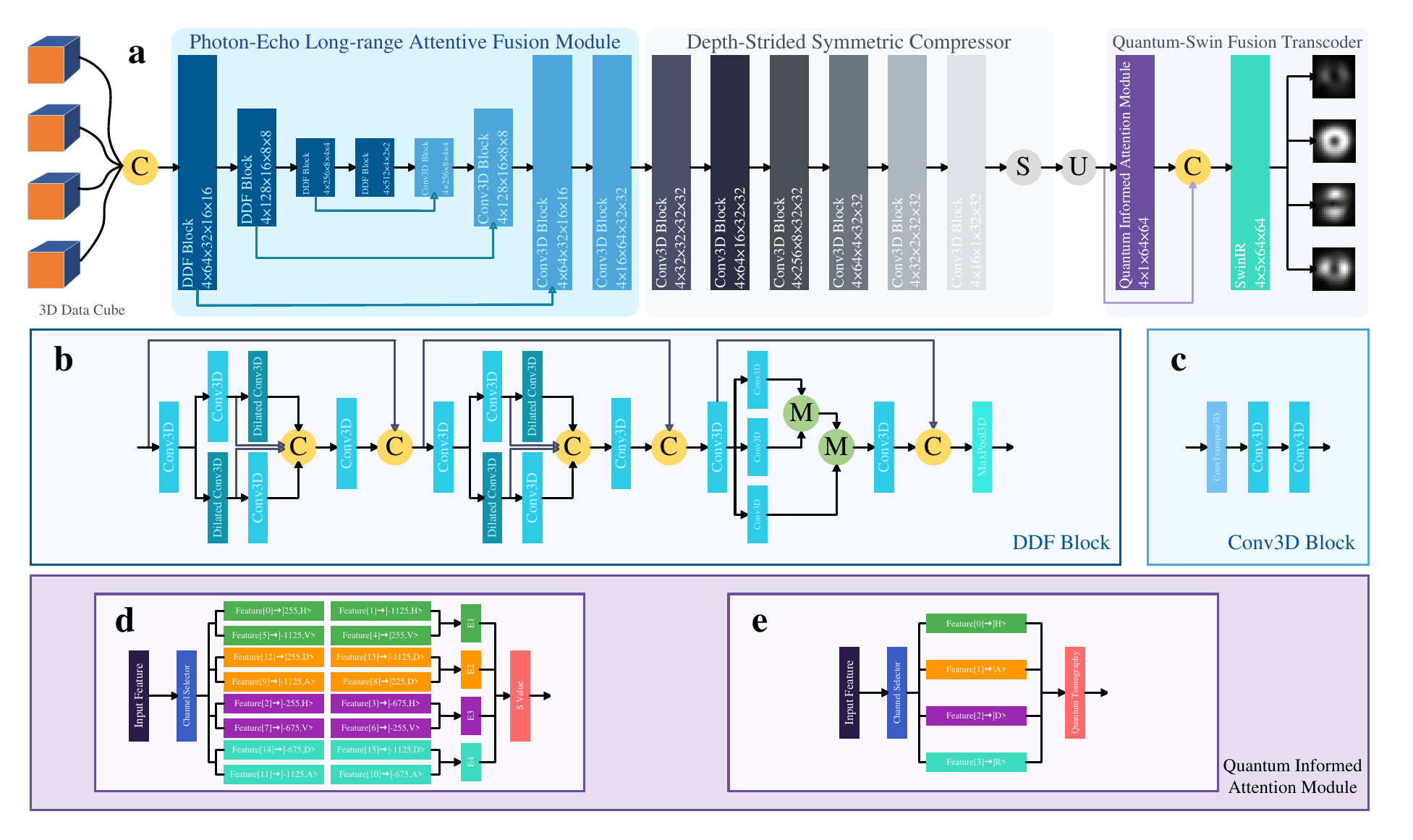} 
\caption{\textbf{Overall architecture of the Quantum-SFNet}. \textbf{a} The computational pipeline processes multiplexed 3D data cubes through three cascaded modules: the Photon-Echo Long-range Attentive Fusion Module, the Depth-Strided Symmetric Compressor, and the Quantum-Swin Fusion Transcoder. Detailed schematic representations are provided for \textbf{b} the multi-branch 3D-DDF block, \textbf{c} the standard Conv3D block, and \textbf{d},\textbf{e} the Quantum Informed Attention Modules  configured for specific quantum measurement tasks.} 
\label{fig:FigureS1} 
\end{figure}

The unified Quantum-SFNet framework (Fig.\ref{fig:FigureS1}a) establishes a computational mechanism for the simultaneous reconstruction of spatio-temporal intensity profiles and the estimation of quantum state parameters, engineered to operate optimally within an intermediate illumination regime. This targeted operational bound systematically circumvents the severe statistical fluctuations characteristic of traditional first-photon imaging while alleviating the experimental necessity of accumulating macroscopic photon ensembles. To accommodate varying quantum measurement protocols, the architecture dynamically concatenates protocol-specific inputs—ranging from 4-channel configurations for quantum state tomography to 16-channel cubes for Bell inequality tests—into unified 5D tensors. These tensors undergo sequential feature abstraction through three cascaded modules: a Photon-Echo Long-range Attentive Fusion Module (PLAFM) that isolates entanglement-governed spatio-temporal correlations, a Depth-Strided Symmetric Compressor (DSC) for hierarchical temporal downsampling , and a Quantum-Swin Fusion Transcoder for parameter extraction  (comprehensive architectural configurations are detailed in subsequent sections). Crucially, quantum-informed attention layers, tailored for metasurface-based measurements, map the extracted cross-dimensional correlations onto basis-specific representations to enable robust single-photon-level pattern recovery. This inference process is rigorously regularized by a physics-constrained hybrid loss function that directly embeds quantum mechanical postulates into the computational model, ensuring that the reconstructed intensity distributions and inferred density matrices strictly adhere to underlying physical laws.

\subsection{Photon-Echo Long-range Attentive Fusion Module}

The robust extraction of entanglement-enhanced spatio-temporal correlations necessitates a hierarchical volumetric learning framework capable of resolving multi-scale physical dependencies. The Photon-Echo Long-range Attentive Fusion Module (PLAFM) utilizes a 3D encoder-decoder architecture, where the encoder systematically compresses spatial dimensions through three progressively contracting blocks (Fig.\ref{fig:FigureS1}b). To adaptively process sparse photon distributions, each block—collectively forming the 3D-DDF module—integrates two multi-branch convolutional units (Fig.\ref{fig:FigureS1}c) followed by a third residual component operating as a Non-Local attention mechanism. Within the multi-branch units, feature extraction proceeds along parallel pathways utilizing standard and dilated $3\times 3\times 3$ convolutions (dilation=2) in reversed sequences. This dual-pathway routing, stabilized by residual connections and $1\times 1\times 1$ channel transformations, simultaneously captures highly localized photon events and broader spatial contexts. The resulting highly compressed latent representations are then globally modeled by a 512-channel Non-Local bottleneck layer, before the decoder reconstructs the spatial domain via transposed convolutions merged with channel-adjusted $1\times 1\times 1$ skip connections. This coupled encoder-decoder configuration, stabilized by Kaiming initialization, effectively aggregates complex photon statistics while maintaining strict computational efficiency.

Extracting long-range dependencies across the spatio-temporal volume relies on correlating non-local signatures embedded within the classical photon counts. The Non-Local self-attention mechanism, functioning as the third residual component within each DDF block, maps the 3D feature tensors using four distinct $1\times 1\times 1$ convolutional transformations. To optimize the high-dimensional tensor operations required by these non-local correlations, the architecture implements grouped attention via systematic channel splitting. Furthermore, to mitigate the gradient vanishing limits endemic to deep spatial compressions, the module incorporates attention weight scaling alongside zero-parameter group normalization (GN), dynamically stabilizing the feature distributions. By mathematically constraining the self-attention operations with these rigorous normalization pathways, the network establishes a stable and physically consistent mechanism for preserving non-local entanglement correlations across successive dimensional transformations.

\subsection{Depth-Strided Symmetric Compressor} 

Efficient mapping of the spatio-temporal volume into the parameter estimation space requires rigorous dimensional reduction. The DSC performs this temporal downsampling through a specialized cascade of strided $3\times 3\times 3$ convolutions. This step systematically projects the high-dimensional outputs from the PLAFM into the compact latent space mandated by the downstream inference modules, stabilized by Kaiming initialization and ReLU-optimized scaling. This deterministic compression separates the essential temporal arrival dynamics required for subsequent state reconstruction while discarding redundant volumetric data.

\subsection{Quantum-Swin Fusion Transcoder} 

The precise extraction of physical observables from the latent representation necessitates an architecture intrinsically adaptable to distinct measurement protocols. The Quantum-swin Fusion Transcoder (QSF) addresses this by coupling a residual-enabled Quantum Informed Attention Module with a SWinIR-based denoising core. For tests of quantum nonlocality (Fig.\ref{fig:FigureS1}d), the network processes spatio-temporal tensors structured as (Batch, 16, Time, Height, Width), extracting states across 16 measurement bases. The framework models Bell inequality violations by evaluating the CHSH parameter:
$\mathcal{S}=|E(a,b)+E(a,b^{\prime})+E(a^{\prime},b)-E(a^{\prime},b^{\prime})|,$
and outputs pixel-resolved $S$-value maps together with the corresponding
16 intensity distributions. Conversely, for quantum state tomography (Fig.\ref{fig:FigureS1}e), the input is dimensionally reparameterized to (Batch, 4, Time, Height, Width) , mapping four measurement bases into unitary evolution operators to yield pixelated density matrices and corresponding intensity profiles. By reconfiguring the attention mechanism to match the basis dimensionality of the specific quantum protocol, the transcoder universally translates abstract statistical features into rigorous quantum mechanical observables.

\subsection{Physics-Constrained Optimization Framework}

Ensuring the physical validity of the network outputs mandates a multi-objective optimization framework strictly bound by quantum mechanical principles. The network parameters are globally optimized utilizing a hybrid loss function defined as:
$$L_{\text{total}}=\alpha L_{\text{matrix}}+\beta L_{\text{intensity}}+\gamma L_{\text{trace}}$$
where $L_{\text{total}}$ represents the overall hybrid objective function governing the network optimization. It is composed of three penalty terms:
$\mathcal{L}_{\mathrm{matrix}}$ denotes the density-matrix element
reconstruction loss, which penalizes deviations between the predicted
and ground-truth matrix elements;
$\mathcal{L}_{\mathrm{intensity}}$ is the intensity-consistency loss,
which encourages the predicted spatial patterns to agree with the
reference data; and
$\mathcal{L}_{\mathrm{trace}}$ is the trace-distance regularization term,
which penalizes deviations from physically valid target states. Utilizing this formulation,  the empirical weighting coefficients $\alpha=1$, $\beta=0.5$, and $\gamma=10^{-4}$ establish a rigorous balance between quantum state preservation and measurement precision. To promote Hermiticity and accurate state reconstruction, the element-wise fidelity loss $L_{\text{matrix}}$ minimizes the Euclidean distance between predicted and ground-truth density matrix components:
$$\begin{aligned}L_{\text{matrix}}&=\frac{1}{4}\left(\left|\left|\rho_{11}^{\text{recon}}-\rho_{11}^{\text{gt}}\right|\right|_2^2+\left|\left|\rho_{22}^{\text{recon}}-\rho_{22}^{\text{gt}}\right|\right|_2^2+\frac{1}{2}\left(\left|\left|\text{Re}(\rho_{12}^{\text{recon}})-\text{Re}(\rho_{12}^{\text{gt}})\right|\right|_2^2\right.\right.\\&\quad\left.\left.+\left|\left|\text{Im}(\rho_{12}^{\text{recon}})-\text{Im}(\rho_{12}^{\text{gt}})\right|\right|_2^2+\left|\left|\text{Re}(\rho_{21}^{\text{recon}})-\text{Re}(\rho_{21}^{\text{gt}})\right|\right|_2^2+\left|\left|\text{Im}(\rho_{21}^{\text{recon}})-\text{Im}(\rho_{21}^{\text{gt}})\right|\right|_2^2\right)\right).\end{aligned}$$
where $\rho_{ij}^{\text{recon}}$ and $\rho_{ij}^{\text{gt}}$ (with $i, j \in \{1, 2\}$) represent the individual elements of the reconstructed and ground-truth $2 \times 2$ density matrices, respectively. Specifically, the diagonal elements, $\rho_{11}$ and $\rho_{22}$, correspond to the real-valued terms of the respective quantum states, while the off-diagonal elements, $\rho_{12}$ and $\rho_{21}$, are complex-valued terms. The operators $\text{Re}(\cdot)$ and $\text{Im}(\cdot)$ extract the real and imaginary components of these off-diagonal terms.

Simultaneously, classical measurement consistency is maintained by the intensity loss:
$$L_{\text{intensity}}=\sum_{\psi\in\{\text{H},\text{V},\text{D},\text{A}\}}\left|\left|\text{I}_\psi^{\text{recon}}-\text{I}_\psi^{\text{gt}}\right|\right|_2^2$$
where $\psi$ denotes projective outcomes in mutually unbiased bases,  $\text{I}_\psi^{\text{recon}}$ represents the reconstructed spatial intensity distribution for a given projection basis $\psi$, while $\text{I}_\psi^{\text{gt}}$ corresponds to the ground-truth intensity profile. This combined objective function explicitly anchors the classical pattern generation to the rigorous mathematical constraints of quantum state representation.

The reconstruction of physically meaningful density matrices requires the predicted states to remain close to valid quantum states. To encourage this behavior, the optimization is further regularized using the trace-distance loss,
\begin{equation*}
\mathcal{L}_{\mathrm{trace}}
=
\frac{1}{2}
\operatorname{Tr}
\left[
\sqrt{
\left(
\rho^{\mathrm{recon}}-\rho^{\mathrm{gt}}
\right)^{\dagger}
\left(
\rho^{\mathrm{recon}}-\rho^{\mathrm{gt}}
\right)
}
\right],
\end{equation*}
where $\rho^{\mathrm{recon}}$ and $\rho^{\mathrm{gt}}$ represent the reconstructed and ground-truth density-matrix operators, respectively, and the superscript $\dagger$ denotes the conjugate transpose. This regularization penalizes deviations of the reconstructed matrices from physically valid target states and thereby reduces the occurrence of nonphysical outputs during training. Combined with the density-matrix element and intensity-consistency losses, it enables the network to jointly reconstruct spatial intensity distributions and estimate quantum-state parameters in a physics-informed manner. The resulting framework therefore functions not merely as an image-processing model, but as a physics-informed estimator whose performance is evaluated through experimental Bell-inequality and quantum-state-tomography tasks.

\subsection{Dataset and Training Strategy}
To achieve physically robust quantum parameter estimation, we construct a hybrid dataset of 576 photon data cubes, covering both noise-free simulated and real experimental samples, and adopt a pretraining and fine-tuning strategy \cite{da2026latent, yin2025rapidly} to decouple the network's representation learning of underlying quantum physical laws from its fitting to hardware systematic errors \cite{wang2026physics, wang2025gradient}. During pretraining, the model is driven by noise-free numerical simulation data, using the Adam optimizer with an initial learning rate of 0.001, embedding quantum-mechanical priors into the latent space of the network backbone. Subsequently, fine-tuning is performed on real experimental data generated by non-overlapping sliding-window sampling along continuous photon event streams. In this stage, we freeze the backbone weights and reduce the learning rate to one-tenth, 0.0001, updating only the parameters of the QSF module at the network's end. This calibrates systematic errors while preventing the model from forgetting the previously learned physical priors. Furthermore, training is terminated when the loss on the validation set, which accounts for 20\% of the samples, reaches saturation, thereby suppressing overfitting to the limited statistical fluctuations of the hardware. The final performance evaluation is conducted on an independent test set comprising 32 experimental photon data cubes, isolated in physical time and without any sliding-window cropping, to avoid data leakage and objectively assess the model's generalization capability.
\section{More Experimental Results}

\subsection{Bell Inequality Violation}

\begin{figure}[htbp]
\centering
\includegraphics[width=1\textwidth]{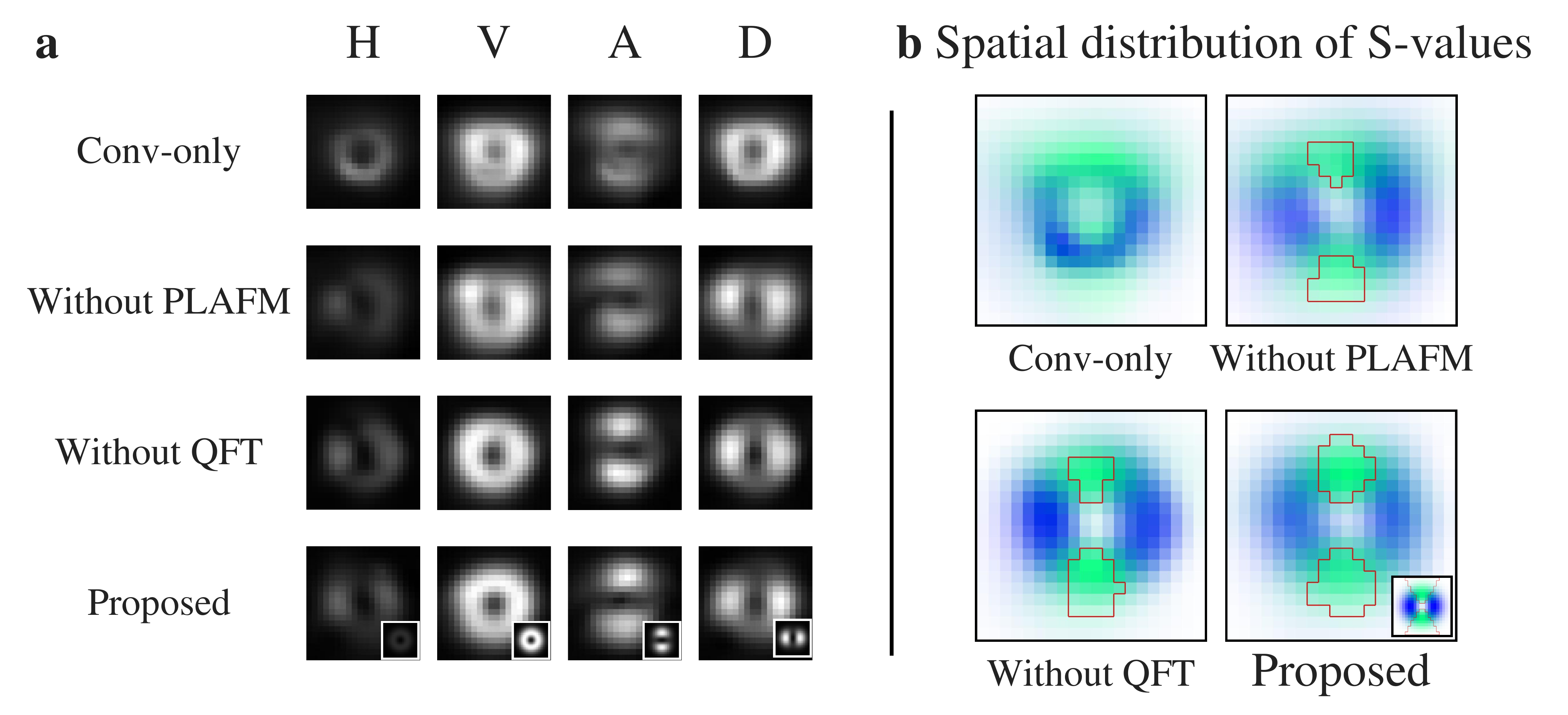}
\caption{\textbf{Experimental verification of Bell inequality violation via the Quantum-SFNet architecture.} \textbf{a} Reconstructed intensity distributions mapped across four measurement bases ($\vert {H}\rangle$, $\vert {V}\rangle$, $\vert {A}\rangle$, and $\vert {D}\rangle$) conditioned on Bob's photon being projected onto the $-22.5^\circ$ linear-polarization state. The comparative panel displays outcomes from the Conv-Only baseline (top row), without PLAFM configuration (second row), without QFT variant (third row), and the complete Quantum-SFNet (bottom row), benchmarked against theoretical numerical simulations (white insets). \textbf{b} Corresponding spatial mappings of the evaluated S-values. Regions demonstrating quantum nonlocality (exceeding the classical limit, $S>2$) are indicated by red boxes, with theoretical predictions provided in the lower-right inset.}
\label{fig:FigureS2}
\end{figure}

The robust extraction of quantum correlations from sparse measurement data necessitates a strictly physics-constrained network architecture. To isolate the physical contribution of individual network components, we present an ablation study comparing the complete Quantum-SFNet against three structural variants: a Conv-Only baseline, a without PLAFM configuration, and a without QFT architecture. This comparative analysis rigorously evaluates the structural dependencies of the model, confirming that the complete, physics-constrained architecture is required to systematically approach the theoretical performance limits of the quantum measurement.

The accurate evaluation of quantum nonlocality demands precise spatial reconstruction of the underlying photon intensity distributions. Figure~\ref{fig:FigureS2}a shows the reconstructed spatial intensity distributions of Alice's photon in the $\ket{H}$, $\ket{V}$, $\ket{A}$, and $\ket{D}$ polarization bases, conditioned on Bob's photon being projected onto the $-22.5^\circ$ linear-polarization state. Quantitative analysis utilizing the Structural Similarity Index (SSIM) (Table \ref{tab:ssim_S}) and Mean Squared Error (MSE) (Table \ref{tab:mse_S}) rigorously differentiates the capabilities of the tested architectures. These metrics confirm that only the complete Quantum-SFNet framework statistically converges to the theoretical physical limits of the expected intensity patterns.

The preservation of underlying quantum entanglement is ultimately verified through the spatially resolved violation of the classical CHSH bound. Computed directly from the reconstructed intensity profiles, the pixelated S-value distributions (Fig. \ref{fig:FigureS2}b) map the localized regions exceeding the classical limit of $S>2$. Operating entirely within classical bounds, the Conv-Only baseline exhibits no nonlocality, while the without PLAFM and without QFT configurations yield only spatially fragmented violations. In contrast, the complete Quantum-SFNet maximizes the observable quantum nonlocality, yielding a $4.28\%$ increase in average S-value magnitude and a $29.41\%$ expansion in spatial coverage relative to the ablated configurations.

\subsection{Effective Qubit Encoding for Imaging Bell-Type Nonlocal Behavior}

To interpret the reconstructed Bell-type violation in a rigorous way, it is essential to specify the effective two-dimensional subspaces and the associated local measurement settings. In our case, the relevant state can be written as
\begin{equation}
\ket{\psi_{\mathrm{exp}}}
=
\frac{1}{\sqrt{2}}
\left(
\ket{H_A,\ell_A=+1}\ket{V_B}
-
\ket{V_A,\ell_A=-1}\ket{H_B}
\right),
\end{equation}
which occupies the composite subspace
\begin{equation}
\mathrm{span}
\left\{
\ket{H_A V_B},
\ket{V_A H_B}
\right\}
\otimes
\mathrm{span}
\left\{
\ket{\ell_A=+1},
\ket{\ell_A=-1}
\right\}.
\end{equation}
Accordingly, we define an effective polarization logical qubit and an OAM qubit as
\begin{equation}
\ket{0}_{p}
\equiv
\ket{H_A V_B},
\qquad
\ket{1}_{p}
\equiv
\ket{V_A H_B},
\end{equation}
and
\begin{equation}
\ket{0}_{o}
\equiv
\ket{\ell_A=+1},
\qquad
\ket{1}_{o}
\equiv
\ket{\ell_A=-1}.
\end{equation}
Within this encoded basis, the state is formally equivalent to a Bell state,
\begin{equation}
\ket{\psi_{\mathrm{exp}}}=\frac{1}{\sqrt{2}}\left(|0\rangle_{p}|0\rangle_{o}-|1\rangle_{p}|1\rangle_{o}\right).
\end{equation}

Under this encoding, the Bell parameter reconstructed from the measured intensity distributions should therefore be interpreted as probing Bell-type nonclassical correlations between the biphoton polarization logical qubit and the OAM qubit of Alice's photon. More precisely, the polarization sector entering the test is not the full four-dimensional two-photon polarization Hilbert space, but only the two-dimensional subspace spanned by $\{|H_AV_B\rangle,|V_AH_B\rangle\}$.

The polarization projections used in the experiment, namely $AA$, $AD$, $LA$, and $LD$, act as measurements in different superposition bases of the logical polarization qubit. Restricting these projections to the subspace $\mathrm{span}\{|H_{A}V_{B}\rangle,|V_{A}H_{B}\rangle\}$, one finds that
\begin{equation}
|AA\rangle \propto \frac{1}{\sqrt{2}}\left(|H_AV_B\rangle+|V_AH_B\rangle\right),
\qquad
|AD\rangle \propto \frac{1}{\sqrt{2}}\left(|H_AV_B\rangle-|V_AH_B\rangle\right),
\end{equation}
which correspond to the $\sigma_x$ basis of the logical qubit, while
\begin{equation}
|LA\rangle \propto \frac{1}{\sqrt{2}}\left(|H_AV_B\rangle-i|V_AH_B\rangle\right),
\qquad
|LD\rangle \propto \frac{1}{\sqrt{2}}\left(|H_AV_B\rangle+i|V_AH_B\rangle\right).
\end{equation}

Therefore, the projection outcomes are identified as
\begin{equation}
a \rightarrow \{LA\},\qquad
a^{\perp} \rightarrow \{LD\},\qquad
a' \rightarrow \{AA\},\qquad
a'^{\perp} \rightarrow \{AD\}.
\end{equation}
The two corresponding dichotomic measurement settings are oriented along the logical Bloch-sphere $y$ and $x$ axes, respectively, up to overall phase conventions and possible relabeling of the $\pm1$ outcomes.

On the OAM side, the local measurement basis is defined in the two-dimensional subspace $\mathrm{span}\{|\ell_A=+1\rangle,|\ell_A=-1\rangle\}$ through azimuthally dependent phase-step projections, which realize states of the form
\begin{equation}
|\theta\rangle_{o}=\frac{1}{\sqrt{2}}\left(|\ell_A=+1\rangle+e^{2i\theta}|\ell_A=-1\rangle\right).
\end{equation}
The corresponding intensity distributions as a function of the azimuthal angle $\theta$ provide the data from which the correlation functions and the global CHSH parameter $S$ can be reconstructed. Consequently, the observed violation should be understood as a Bell-type violation in an effective hybrid two-qubit space formed by the biphoton polarization logical qubit and the OAM qubit. 

To explicitly formulate the Bell inequality test, we designate the two measurement settings for the OAM qubit as $b$ and $b'$, corresponding to the specific azimuthal projection angles $\theta = 22.5^\circ$ and $\theta' = 67.5^\circ$, respectively. Combined with the logical polarization settings $a$ and $a'$, this specific configuration enables the evaluation of the CHSH polynomial. The observed violation should therefore be interpreted as evidence of Bell-type nonclassical correlations within the effective hybrid two-qubit space formed by the biphoton polarization logical qubit and the spatial OAM qubit of Alice's photon.

\subsection{Quantum State Tomography in Remote State Preparation}

\begin{figure}[htbp]
\centering
\includegraphics[width=1\textwidth]{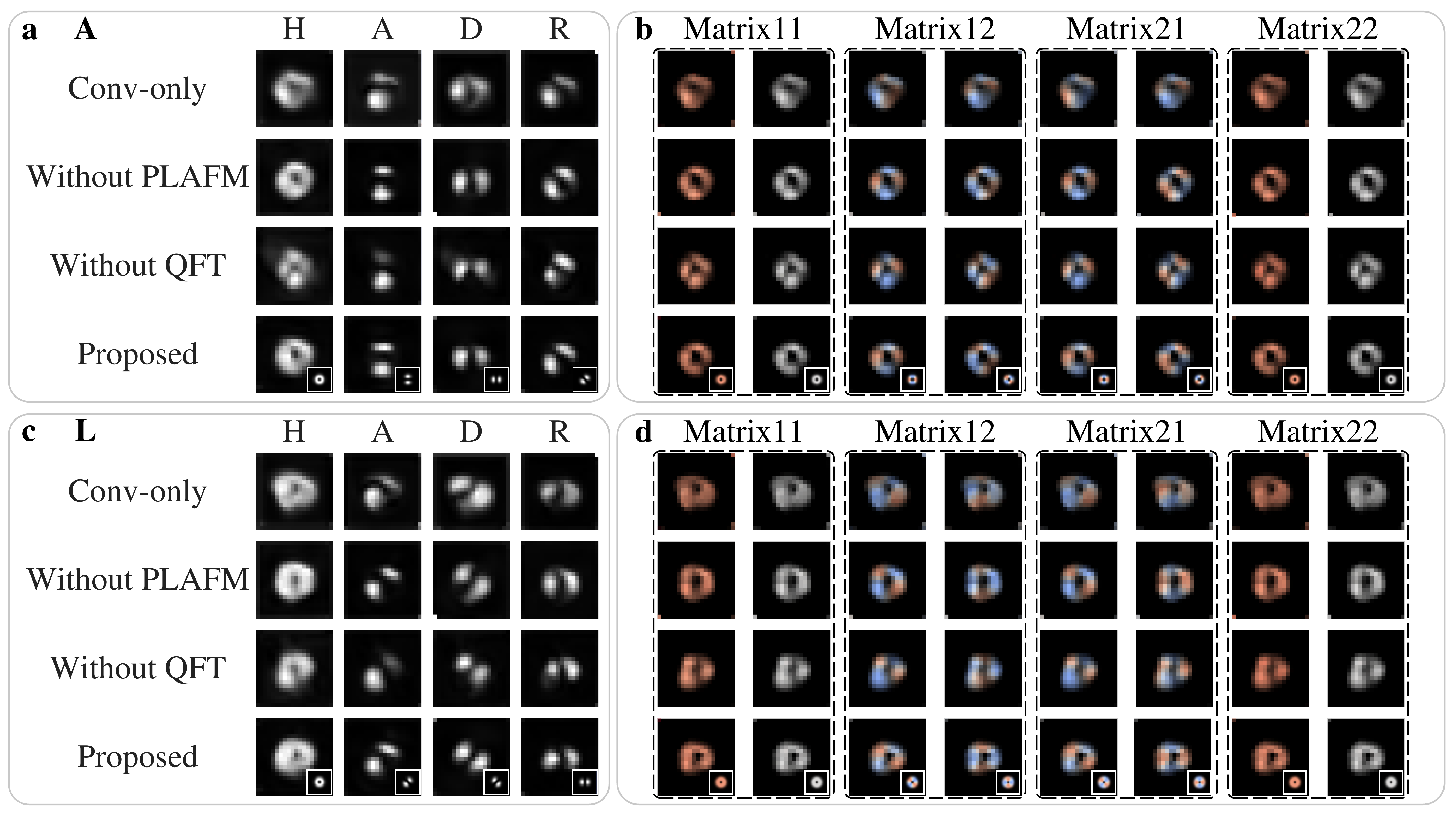}
\caption{\textbf{Experimental quantum state tomography via the Quantum-SFNet architecture.}
\textbf{a}, \textbf{c}, Reconstructed spatial intensity distributions of Alice's photon in four polarization projection states
($\ket{H}$, $\ket{A}$, $\ket{D}$, and $\ket{R}$),
conditioned on Bob's photon being projected onto
\textbf{a} the $\ket{A}$ state and
\textbf{c} the $\ket{L}$ state, respectively.
The comparative panels display results from the Conv-Only baseline (top row), the without PLAFM configuration (second row), the without QFT variant (third row), and the complete Quantum-SFNet (bottom row), benchmarked against theoretical numerical simulations shown in the white insets.
\textbf{b}, \textbf{d}, Corresponding reconstructed density matrices of Alice's conditionally prepared photon, separated into real and imaginary components and derived from the intensity distributions in \textbf{a} and \textbf{c}, respectively, with numerical simulations shown in the bottom-row insets.}
\label{fig:FigureS3}
\end{figure}

Comprehensive characterization of the remotely prepared quantum states necessitates full quantum state tomography. By leveraging the network's adaptive capacity, the measurement sequence is reconfigured through the substitution of the CHSH module with a tomographic estimator within the Quantum Informed Attention Modules. Figure~\ref{fig:FigureS3}a shows the reconstructed spatial intensity distributions of Alice's photon in the $\ket{H}$, $\ket{A}$, $\ket{D}$, and $\ket{R}$ polarization  projection states, conditioned on Bob's photon being projected onto the $\ket{A}$ state. Across the evaluated architectures, the Conv-Only variant yields incomplete petal structures and an inability to resolve the expected circular geometry of the $\vert {H}\rangle$ state. Similarly, the without QFT implementation introduces anomalous intensity modulations in the $\vert {A}\rangle$ state. In contrast, the complete Quantum-SFNet reproduces the spatial modes of the numerical simulations (white insets). These visual assessments are rigorously corroborated by MSE and SSIM metrics (Tables \ref{tab:MSE_A} and \ref{tab:ISSIM_A}).

The physical accuracy of the quantum state characterization is strictly bounded by the fidelity of the preceding intensity reconstruction. The computed density matrices (Fig. \ref{fig:FigureS3}b), delineated into real and imaginary components, directly reflect the architectural variations observed in the spatial intensity profiles. Evaluation of the derived matrices utilizing trace distance (Table \ref{tab:trace_distance_A}) and density matrix SSIM (Table \ref{tab:RSSIM_A}) systematically distinguishes the performance of each network configuration. These metrics establish that the complete framework produces a statistically superior density matrix reconstruction, converging strictly to the theoretical quantum state representation.

To evaluate the robustness of the tomographic reconstruction under different remote projection conditions, Figs.~\ref{fig:FigureS3}c and \ref{fig:FigureS3}d present the reconstructed intensity distributions and corresponding density matrices of Alice's photon conditioned on Bob's photon being projected onto the $\ket{L}$ state. The corresponding quantitative analysis (Tables \ref{tab:MSE_L}-\ref{tab:RSSIM_L}) aggregates MSE and SSIM for the intensity profiles alongside trace distance and matrix SSIM for the density operators. This consistent performance across distinct measurement settings demonstrates the generalized stability of the physical parameter estimation.

Ultimately, the accurate mapping of sparse photon statistics to rigorous quantum observables requires the strict imposition of theoretical constraints during the inference process. The comprehensive evaluation, integrating localized spatial pattern resolution with global quantum state metrics, confirms the dual functionality of the architecture in classical intensity recovery and quantum parameter estimation. The measurable performance divergence between the complete quantum-enhanced framework and the without QFT implementation explicitly quantifies the contribution of the physics-constrained topological module. Embedding physical measurement postulates directly into the neural architecture is thus demonstrated to be a critical requirement for achieving high-fidelity quantum state tomography from highly sparse data regimes.

\section{Quantitative Evaluation}

\subsection{Experimental Intensity Reconstruction Metrics of Bell's Inequality}

The accurate preservation of spatial photon correlations requires high-fidelity reconstruction of the underlying intensity modes. Table \ref{tab:ssim_S} quantifies this spatial fidelity utilizing the SSIM across four distinct measurement bases ($-112.5^{\circ}$, $-67.5^{\circ}$, $-22.5^{\circ}$, and $+22.5^{\circ}$). The proposed Quantum-SFNet achieves the best overall SSIM performance across the four measurement bases, attaining the highest value in three cases and matching the best-performing variant in the remaining case. These results indicate that the proposed physics-informed architecture improves the overall preservation of the spatial mode structure.

\begin{table}[htbp]
\centering
\caption{\textbf{Structural Similarity Index (SSIM) of the reconstructed intensity distributions corresponding to Fig. \ref{fig:FigureS2}a}. The metrics benchmark the Conv-Only baseline, the without PLAFM configuration, the without QFT variant, and the proposed Quantum-SFNet against theoretical numerical simulations, quantifying the spatial fidelity of the reconstructed photon modes.}
\label{tab:ssim_S}
\begin{tabular}{l *{4}{S[table-format=1.4]}}
\toprule
\textbf{Measurement basis} & \textbf{Conv-Only} & \textbf{without PLAFM} & \textbf{without QFT} & \textbf{Proposed} \\
\midrule
$-112.5^{\circ}$ & 0.1980 & 0.2044 & 0.2304 & \textbf{0.2305} \\
$-67.5^{\circ}$ & 0.2028 & 0.1977 & 0.2044 & \textbf{0.2266} \\
$-22.5^{\circ}$ & 0.1955 & 0.2099 & \textbf{0.2238} & \textbf{0.2238} \\
$+22.5^{\circ}$ & 0.1803 & 0.1812 & 0.1801 & \textbf{0.1987} \\
\bottomrule
\end{tabular}
\end{table}

Minimizing absolute statistical deviations from theoretical predictions is essential for the reliable extraction of quantum parameters. As detailed in Table \ref{tab:mse_S}, the MSE provides a strict bound on the amplitude reconstruction accuracy relative to the numerical simulations. The complete framework systematically suppresses these reconstruction errors, achieving an MSE as low as $0.0055$ in the $+22.5^{\circ}$ basis, thereby outperforming the unconstrained Conv-Only baseline, which exhibits statistical fluctuations nearly an order of magnitude higher. By actively minimizing these amplitude discrepancies, the physics-constrained model ensures that the derived intensity distributions converge stably to the fundamental physical observables required for rigorous Bell inequality evaluation.

\begin{table}[htbp]
\centering
\caption{\textbf{Mean Squared Error (MSE) analysis of the reconstructed intensity profiles presented in Fig. \ref{fig:FigureS2}a}. The evaluation compares the statistical deviations of the four architectural variants from the predicted numerical limits across the specified measurement bases.}
\label{tab:mse_S}
\begin{tabular}{l *{4}{S[table-format=1.4]}}
\toprule
\textbf{Measurement basis} & \textbf{Conv-Only} & \textbf{without PLAFM} & \textbf{without QFT} & \textbf{Proposed} \\
\midrule
$-112.5^{\circ}$ & 0.0301 & 0.0342 & 0.0110 & \textbf{0.0076} \\
$-67.5^{\circ}$ & 0.0200 & 0.0268 & 0.0113 & \textbf{0.0093} \\
$-22.5^{\circ}$ & 0.0357 & 0.0348 & \textbf{0.0083} & 0.0085 \\
$+22.5^{\circ}$ & 0.0467 & 0.0430 & 0.0066 & \textbf{0.0055} \\
\bottomrule
\end{tabular}
\end{table}

\subsection{Tomographic Reconstruction Metrics}

Accurate tomographic reconstruction relies fundamentally on minimizing the statistical variance between measured and theoretical intensity distributions. Table~\ref{tab:MSE_A} quantifies this variance using the MSE of the reconstructed intensity profiles of Alice's photon in the $\ket{H}$, $\ket{A}$, $\ket{D}$, and $\ket{R}$ polarization bases, conditioned on Bob's photon being projected onto the $\ket{A}$ state. The proposed Quantum-SFNet consistently suppresses amplitude deviations, achieving minimum MSE values across the bases (e.g., $0.0010$ for $\vert {D}\rangle$), while the Conv-Only and without QFT architectures exhibit significantly larger statistical fluctuations. This suppression of amplitude error indicates that the complete physics-constrained architecture effectively regularizes the inverse problem, yielding classical intensity profiles strictly consistent with the underlying quantum state.

\begin{table}[htbp]
\centering
\caption{\textbf{MSE of the reconstructed intensity profiles of Alice's photon conditioned on Bob's photon being projected onto the $\ket{A}$ state, corresponding to Fig.~\ref{fig:FigureS3}a}. The evaluation compares the statistical deviations of the Conv-Only baseline, the without PLAFM configuration, the without QFT variant, and the proposed network. Lower MSE values indicate higher amplitude accuracy relative to the theoretical numerical results.}
\label{tab:MSE_A}
\begin{tabular}{l *{4}{S[table-format=1.4]}}
\toprule
\textbf{Architecture} & $\vert {H}\rangle$ & $\vert {A}\rangle$ & $\vert {D}\rangle$ & $\vert {R}\rangle$ \\
\midrule
Conv-Only & 0.0026552 & 0.024975 & 0.0089862 & 0.0100520 \\
without PLAFM & 0.0035941 & 0.0033188 & 0.0031328 & 0.0040730 \\
without QFT & 0.0050471 & 0.0080206 & 0.0030525 & 0.0039566 \\
Proposed & \textbf{0.0019948} & \textbf{0.0019916} & \textbf{0.0010534} & \textbf{0.0016491} \\
\bottomrule
\end{tabular}
\end{table}

Beyond absolute amplitude precision, the topological fidelity of the spatial modes dictates the accuracy of the extracted phase information. As evaluated by the SSIM in Table~\ref{tab:ISSIM_A}, the spatial correlation of the reconstructed patterns of Alice's photon is benchmarked against the numerical ground truth, conditioned on Bob's photon being projected onto the $\ket{A}$ state. The complete framework provides the best overall structural overlap, yielding the highest SSIM values in the $\vert H\rangle$,
$\vert D\rangle$, and $\vert R\rangle$ bases, while the without PLAFM
variant performs slightly better in the $\vert A\rangle$ basis.
Overall, the proposed architecture demonstrates robust preservation
of the spatial mode structure across the evaluated measurement bases. Maintaining these spatial correlations confirms the necessity of the proposed attention mechanisms in preserving the geometric phase signatures essential for high-fidelity state inversion.

\begin{table}[htbp]
\centering
\caption{\textbf{SSIM of the reconstructed intensity profiles of Alice's photon conditioned on Bob's photon being projected onto the $\ket{A}$ state, corresponding to Fig.~\ref{fig:FigureS3}a}. Higher SSIM values indicate better preservation of the spatial modes relative to the ground-truth numerical simulations.}
\label{tab:ISSIM_A}
\begin{tabular}{l *{4}{S[table-format=1.4]}}
\toprule
\textbf{Architecture} & $\vert {H}\rangle$ & $\vert {A}\rangle$ & $\vert {D}\rangle$ & $\vert {R}\rangle$ \\
\midrule
Conv-Only & 0.42051 & 0.059901 & 0.36469 & 0.35925 \\
without PLAFM & 0.42625 & \textbf{0.48024} & 0.31775 & 0.34735 \\
without QFT & 0.29258 & 0.31112 & 0.31062 & 0.35558 \\
Proposed & \textbf{0.54868} & 0.44753 & \textbf{0.41678} & \textbf{0.4679} \\
\bottomrule
\end{tabular}
\end{table}

The ultimate validity of the tomographic process is quantified by the geometric proximity of the reconstructed density matrix to the theoretical ideal state. Table~\ref{tab:trace_distance_A} presents the trace distance between the reconstructed and theoretical states of Alice's photon conditioned on Bob's photon being projected onto the $\ket{A}$ state. The complete architecture minimizes this metric to $0.1018$, demonstrating a substantial convergence improvement over the Conv-Only baseline ($0.3166$) and the without QFT variant ($0.1889$). By approaching the theoretical limit of zero trace distance, the full network ensures that the computationally inferred density operator represents a physically valid and highly pure quantum state.

\begin{table}[htbp]
\centering
\caption{\textbf{Trace distance of the reconstructed density matrices of Alice's photon conditioned on Bob's photon being projected onto the $\ket{A}$ state, corresponding to Fig.~\ref{fig:FigureS3}b}. Lower values indicate closer agreement between the reconstructed density matrices and the theoretical target states.}
\label{tab:trace_distance_A}
\begin{tabular}{l *{1}{S[table-format=1.4]}}
\toprule
\textbf{Architecture} & {trace distance} \\
\midrule
Conv-Only & 0.31667 \\
without PLAFM & 0.10830 \\
without QFT & 0.18890 \\
Proposed & \textbf{0.10180} \\
\bottomrule
\end{tabular}
\end{table}

The precise reconstruction of quantum coherence necessitates accurate estimation of both the real and imaginary off-diagonal density-matrix elements. Table~\ref{tab:RSSIM_A} reports the component-wise SSIM of the reconstructed density matrices of Alice's photon conditioned on Bob's photon being projected onto the $\ket{A}$ state. The proposed model achieves superior fidelity across the matrix elements, notably stabilizing the sensitive imaginary coherences, $\operatorname{Im}(\rho_{12})$ and $\operatorname{Im}(\rho_{21})$, near $0.70$, whereas the without PLAFM and without QFT configurations exhibit appreciable degradation. This uniform component-wise accuracy demonstrates that the network reliably preserves the phase coherence of the remotely prepared state of Alice's photon, avoiding unphysical dephasing artifacts during reconstruction.

\begin{table}[htbp]
\centering
\caption{\textbf{Component-wise SSIM of the reconstructed density matrices of Alice's photon conditioned on Bob's photon being projected onto the $\ket{A}$ state, corresponding to Fig.~\ref{fig:FigureS3}b}. The evaluation separately considers the real and imaginary components of the density matrices to quantify the preservation of quantum coherence.}
\label{tab:RSSIM_A}
\begin{tabular}{l *{6}{S[table-format=1.4]}}
\toprule
\textbf{Architecture} & {$\rho_{11}$} & {$\text{Re}(\rho_{12})$} & {$\text{Im}(\rho_{12})$} & {$\text{Re}(\rho_{21})$} & {$\text{Im}(\rho_{21})$} & {$\rho_{22}$} \\
\midrule
Conv-Only & 0.92073 & 0.21097 & 0.17377 & 0.172 & 0.1789 & 0.94064\\
without PLAFM & 0.9024 & 0.633 & 0.69017 & 0.63838 & 0.69332 & 0.90722 \\
without QFT & 0.79223 & 0.46272 & 0.31939 & 0.45437 & 0.32188 & 0.77364 \\
Proposed & \textbf{0.9303} & \textbf{0.69056} & \textbf{0.699} & \textbf{0.69106} &\textbf{0.70015} & \textbf{0.95007}\\
\bottomrule
\end{tabular}
\end{table}

To evaluate the robustness of the tomographic inversion under a different remote projection condition, Table~\ref{tab:MSE_L} extends the MSE analysis to the reconstructed intensity profiles of Alice's photon conditioned on Bob's photon being projected onto the $\ket{L}$ state. The proposed architecture maintains its low-error performance, particularly in the $\vert {A}\rangle$ and $\vert {R}\rangle$ bases, while the baseline models continue to show basis-dependent statistical vulnerabilities. This consistent amplitude recovery under different remote projection conditions demonstrates the generalized adaptability of the network's measurement-to-state mapping.

\begin{table}[htbp]
\centering
\caption{\textbf{MSE of the reconstructed intensity profiles of Alice's photon conditioned on Bob's photon being projected onto the $\ket{L}$ state, corresponding to Fig.~\ref{fig:FigureS3}c}.}
\label{tab:MSE_L}
\begin{tabular}{l *{4}{S[table-format=1.4]}}
\toprule
\textbf{Architecture} & $\vert {H}\rangle$ & $\vert {A}\rangle$ & $\vert {D}\rangle$ & $\vert {R}\rangle$ \\
\midrule
Conv-Only & 0.0049708 & 0.013425 & 0.018875 & 0.01134 \\
without PLAFM & 0.0052021 & 0.0035666 & \textbf{0.0089511} & 0.0053936 \\
without QFT & \textbf{0.0040881} & 0.011405 & 0.011591 & 0.0043132 \\
Proposed & 0.0040959 & \textbf{0.0021954} & 0.01019 & \textbf{0.0017738} \\
\bottomrule
\end{tabular}
\end{table}

Consistent spatial-mode preservation under different remote projection conditions is required to avoid systematic bias in parameter estimation. Table~\ref{tab:ISSIM_L} assesses the structural fidelity of the reconstructed intensity profiles of Alice's photon conditioned on Bob's photon being projected onto the $\ket{L}$ state. The proposed Quantum-SFNet continues to yield the highest structural correlations, peaking at $0.6227$ for the $\vert {R}\rangle$ state, effectively mitigating the mode distortion present in the ablated networks. The sustained spatial fidelity confirms that the network's topological feature extraction operates independently of the specific input state phase.

\begin{table}[htbp]
\centering
\caption{\textbf{SSIM of the reconstructed intensity profiles of Alice's photon conditioned on Bob's photon being projected onto the $\ket{L}$ state, corresponding to Fig.~\ref{fig:FigureS3}c}.}
\label{tab:ISSIM_L}
\begin{tabular}{l *{4}{S[table-format=1.4]}}
\toprule
\textbf{Architecture} & $\vert {H}\rangle$ & $\vert {A}\rangle$ & $\vert {D}\rangle$ & $\vert {R}\rangle$ \\
\midrule
Conv-Only & 0.35806 & 0.11623 & 0.29548 & 0.3385 \\
without PLAFM & 0.45759 & 0.42704 & 0.28043 & 0.28281 \\
without QFT & 0.38889 & 0.33148 & 0.25824 & 0.39391 \\
Proposed & \textbf{0.51401} & \textbf{0.43194} & \textbf{0.39077} & \textbf{0.62277} \\
\bottomrule
\end{tabular}
\end{table}

A rigorous tomographic estimator must yield physical density operators under different remote projection conditions. As shown in Table~\ref{tab:trace_distance_L}, the complete network reconstructs the state of Alice's photon conditioned on Bob's photon being projected onto the $\ket{L}$ state with a trace distance of $0.1024$, effectively halving the value obtained using the without QFT variant ($0.2285$). This persistent minimization of state distinguishability affirms the universal applicability of the physics-constrained architecture in robust quantum state characterization.

\begin{table}[htbp]
\centering
\caption{\textbf{Trace distance of the reconstructed density matrices of Alice's photon conditioned on Bob's photon being projected onto the $\ket{L}$ state, corresponding to Fig.~\ref{fig:FigureS3}d}. Lower values indicate closer agreement between the reconstructed density matrices and the theoretical target states.}
\label{tab:trace_distance_L}
\begin{tabular}{l *{1}{S[table-format=1.4]}}
\toprule
\textbf{Architecture} & {trace distance} \\
\midrule
Conv-Only & 0.28756 \\
without PLAFM & 0.12341 \\
without QFT & 0.22851 \\
Proposed & \textbf{0.10242} \\
\bottomrule
\end{tabular}
\end{table}

The stability of the inferred quantum coherence under different remote projection conditions further demonstrates the robustness of the computational model. Table \ref{tab:RSSIM_L} tracks the matrix component SSIM for the $\vert {L}\rangle$ state reconstruction. The full architecture achieves the best overall component-wise SSIM,
yielding the highest values for five of the six matrix components and
remaining comparable to the without PLAFM variant for
$\operatorname{Re}(\rho_{21})$. In particular, it provides the best
reconstruction of both imaginary coherence components, demonstrating
robust recovery of the phase-dependent quantum-state information. The ability to reliably invert phase-dependent coherences across multiple bases confirms the rigorous integration of physical measurement postulates within the network topology.

\begin{table}[htbp]
\centering
\caption{\textbf{Component-wise SSIM of the reconstructed density matrices of Alice's photon conditioned on Bob's photon being projected onto the $\ket{L}$ state, corresponding to Fig.~\ref{fig:FigureS3}d}.}
\label{tab:RSSIM_L}
\begin{tabular}{l *{6}{S[table-format=1.4]}}
\toprule
\textbf{Architecture} & {$\rho_{11}$} & {$\text{Re}(\rho_{12})$} & {$\text{Im}(\rho_{12})$} & {$\text{Re}(\rho_{21})$} & {$\text{Im}(\rho_{21})$} & {$\rho_{22}$} \\
\midrule
Conv-Only & 0.91231 & 0.17956 & 0.29582 & 0.16494 & 0.31688 & 0.93653\\
without PLAFM & 0.91294 & 0.68685 & 0.57193 & \textbf{0.68739} & 0.57337 & 0.9183 \\
without QFT & 0.81653 & 0.32499 & 0.29606 & 0.31653 & 0.29945 & 0.78431 \\
Proposed & \textbf{0.94334} & \textbf{0.69154} & \textbf{0.70897} & 0.68532 &\textbf{0.71287} & \textbf{0.95005}\\
\bottomrule
\end{tabular}
\end{table}
\newpage
\bibliography{Supplementary_Materials/suppref}